\documentclass[a4paper,14pt]{article}
\usepackage{jheppub} 
\usepackage{lineno}
\usepackage{booktabs}
\usepackage[normalem]{ulem}
\usepackage{color,graphicx,float,subfigure,xcolor}
\usepackage{amsfonts,amssymb,theorem,mathrsfs,times}
\usepackage{bm}
\usepackage{mathtools}
\usepackage{amsfonts,amssymb,theorem,mathrsfs}
\usepackage{dsfont}
\usepackage{setspace}
\usepackage{amsmath}  

\usepackage{accents}
\usepackage{url}

\arxivnumber{2606.22580} 

\title{\boldmath  Extended parameterized spin expansion formalism for ringdown analysis with GW250114}

\author[a,b,c]{Jia-Ning Chen}
\author[a]{Liang-Bi Wu}
\author[b,c,a]{Zong-Kuan Guo}

\affiliation[a]{School of Fundamental Physics and Mathematical Sciences, Hangzhou Institute for Advanced Study, UCAS, Hangzhou 310024, China}

\affiliation[b]{CAS Key Laboratory of Theoretical Physics, Institute of Theoretical Physics, Chinese Academy of Sciences, Beijing 100190, China}
 
\affiliation[c]{University of Chinese Academy of Sciences, Beijing 100190, China}

\emailAdd{chenjianing22@mails.ucas.ac.cn}
\emailAdd{wulb@ucas.ac.cn}
\emailAdd{guozk@itp.ac.cn}

\abstract{Parameterized descriptions of black-hole quasinormal-mode spectra are essential for testing gravity with ringdown observations. We extend the Parameterized Spin Expansion Coefficients (ParSpec) formalism by simultaneously sampling the characteristic length scale $\tilde{\ell}$ and the scaling index $\tilde{p}$, rather than fixing $\tilde{p}$ to a theory-motivated integer and constraining $\tilde{\ell}$. Physically, this extension promotes the scaling of the spectral corrections coming from higher-curvature operators to an observable quantity. Methodologically, it enables us to investigate the enlarged ParSpec parameter space and identify the prior geometry induced by conditions on the effective coupling $\gamma$. We examine the robustness of the framework by using the $220$ and $220+221$ ringdown models over different start times with informative priors on mass and luminosity for GW250114, and further study a joint constraint with GW231123. We pointed out that $\gamma<1$ avoids an artificial correlation between $\tilde{p}$ and $\tilde{\ell}$ among the coupling prescriptions considered, and the choices of ringdown starting time and QNM mode content in the ParSpec framework should be matched to the information content and resolving power of the data. Higher-SNR ringdowns and hierarchical analyses of a larger event population will be required to break the $\tilde{p}$-$\tilde{\ell}$ degeneracy and directly probe the scaling structure of corrections to GR.}

\begin{document}
\maketitle
\flushbottom

\section{Introduction}

The ringdown signal of a binary black-hole merger is governed by the quasinormal-modes (QNMs) spectrum of the remnant black hole~\cite{Nollert:1999ji,Berti:2009kk,Konoplya:2011qq}. Measuring this signal with gravitational-wave detectors~\cite{LIGOScientific:2016aoc,LIGOScientific:2014pky,KAGRA:2013rdx} therefore provides a direct way to perform black-hole spectroscopy~\cite{Berti:2025hly}. In general relativity (GR), the remnant is expected to relax to a Kerr black hole, whose QNM frequencies and damping times are uniquely determined by the remnant mass and spin~\cite{Teukolsky:1972my,Teukolsky:1973ha, Leaver:1985ax,Onozawa:1996ux, Berti:2005ys,Yang:2012he, Stein:2019mop}. In theories beyond GR~\cite{Alexander:2009tp,Sotiriou:2013qea, Clifton:2011jh,Chen:2025wfi}, by contrast, the QNM spectra can depend on additional degrees of freedom, leading to possible deviations from the Kerr case. Therefore, it is essential for testing gravity theories to develop flexible and computationally efficient parametrized frameworks that can capture such deviations while remaining compatible with ringdown data analyses~\cite{Berti:2018vdi}.

A standard way to search for deviations from the Kerr spectra is to introduce phenomenological shifts in the QNM frequencies and damping times, such as $\delta f_{\ell mn}$ and $\delta\tau_{\ell mn}$~\cite{Gossan:2011ha, Meidam:2014jpa,Carullo:2018sfu}. This approach is useful for testing the consistency of individual events with GR~\cite{Isi:2019aib, LIGOScientific:2020tif}. A more efficient strategy is to parametrize the QNM spectrum in a structured form, and to relate source-independent coefficients that can be constrained jointly using multiple gravitational-wave events. The Parameterized Spin Expansion Coefficients (ParSpec) formalism provides such a framework~\cite{Maselli:2019mjd,Carullo:2021dui}. In this framework, the QNM frequency and damping time are expanded as polynomial functions of the dimensionless remnant spin, while the dimensionless coupling parameter $\gamma=\left(\frac{\ell c^2}{G M_{\rm s}}\right)^p$ is introduced, where $\ell$ is a characteristic length scale coming from the effective-field theory~(EFT) and is treated as a source-independent parameter. This approach provides a natural way to interpret the derivations from GR in ringdown measurements within the EFT framework, and constrains the fundamental length scale through joint multi-event observations. Previous studies have typically fixed the scaling index $p$ to an integer~\cite{Carullo:2021dui, Silva:2022srr, Maenaut:2024oci, Chung:2025wbg}, corresponding to a specific modified-gravity theory or EFT operator, and then performed joint constraints on the characteristic length scale $\ell$ across multiple events.

Additionally, the ParSpec formalism has also been incorporated into \texttt{pyRing}, a Bayesian inference package designed for time-domain ringdown analyses~\cite{Carullo:2019flw}. Along with \texttt{ringdown}~\cite{Isi:2019aib, Isi:2021iql}, the two analysis tools, as the infrastructure for gravitational-wave ringdown investigations, improve the efficiency of parameter estimation and accelerate observational tests of black-hole spectroscopy. During the O1-O4 observing runs~\cite{LIGOScientific:2018mvr, LIGOScientific:2020ibl, LIGOScientific:2021usb, KAGRA:2021vkt, LIGOScientific:2025slb, LIGOScientific:2026wfs}, the LIGO-Virgo-KAGRA~(LVK) Collaboration has detected more than three hundred GW events, whose ringdown signals provide valuable probes of black-hole properties and strong-field gravity. The upcoming O5 run is expected to mark a new stage in GW astronomy~\cite{LIGOScientific:2026sit}, as a larger population of high-SNR events will enable more precise ringdown analyses and deeper tests of the nature of gravity.

In this work, we dispose of the assumption that $p$ is an integer~\cite{Carullo:2021dui}, and extend the ParSpec formalism by simultaneously sampling the scaling index $p$ and the characteristic length scale $\ell$, which are denoted by $\tilde{p}$ and $\tilde{\ell}$ in subsequent analysis, respectively. There are two motivations for relaxing the assumption and treating it as a continuously sampled parameter. Phenomenologically, real data may not correspond exactly to a single clean higher-curvature operator. Effects such as operator mixing, truncation errors~\cite{Furnstahl:2015rha,Endlich:2017tqa,Sennett:2019bpc}, or even environmental effects~\cite{Barausse:2014tra,Cardoso:2021wlq} may lead to an effective non-integer value of $\tilde{p}$. In this case, treating $\tilde{p}$ as a continuous parameter in the ParSpec method can absorb such complex behavior. If the posterior of $\tilde{p}$ favors an integer value, the data support a scale dependence similar to that of a higher-curvature operator; if the posterior favors a non-integer value, the deviation is difficult to interpret as the effect of a single higher-curvature, suggesting richer phenomenology. In this sense, this approach can be viewed as a theory-agnostic test.

Methodologically, this extension allows us to study the original ParSpec parameterization in a broader parameter space. While previous works usually fixed $\tilde{p}$ and constrained $\tilde{\ell}$ under the constant condition on the effective coupling, which is $\gamma<1$, we investigate their posterior distributions when both are sampled simultaneously under different conditions on $\gamma$. With the loudest event GW250114~\cite{LIGOScientific:2025wao,LIGOScientific:2025rid}, we further use different ringdown start times and QNMs models, including the $220$-only and $220+221$ cases, to assess the robustness of the method. Although a degeneracy between $\tilde{p}$ and $\tilde{\ell}$ is expected, this extension provides a useful complement to the original ParSpec framework. Informative priors on mass and luminosity are adopted in the Bayesian sampling in order to mitigate parameter degeneracies and improve the stability of the inference. A joint analysis including GW231123~\cite{LIGOScientific:2025rsn} is also performed to examine the behavior of the framework in a multi-event setting.

The remaining part of this paper is organized as follows. In Sec. \ref{sec:method}, we introduce the extended ParSpec and describe the Bayesian setup for ringdown analysis. In Sec. \ref{sec: Results and Discussions}, we present the results of the posterior distribution, discuss the role of the coupling condition, compare the $220$-only and $220+221$ analyses, and quantify the information gain using Kullback-Leibler (KL) divergences. We also construct the posterior marginalized over the ringdown start time using different weights. Sec. \ref{sec: Conclusion} summarizes our conclusions. Additional results for different priors on $\tilde{p}$ and for the inclusion of GW231123 are presented in the Appendix \ref{apx:the results of the prior choice} and Appendix \ref{appendix:GW231123}, respectively.

\section{Method}
\label{sec:method}
Except for introducing the new polarization modes of GW, the most direct way of a black hole beyond GR affecting the ringdown phase is to incorporate the corrections to the QNM spectra of a Kerr BH $\bar{\omega}_{\mathrm{GR}}$, resulting in the spectra beyond GR $\bar{\omega}_{\mathrm{non-GR}}$. The spectra $\bar{\omega}=2 \pi f + \mathrm{i} /\tau$ can be formally written as 
\begin{eqnarray}
    f_{\mathrm{non-GR}} & =& f_{\mathrm{GR}} + \delta f\, , \nonumber\\
       \tau_{\mathrm{non-GR}}  & = &\tau_{\mathrm{GR}} + \delta \tau \, ,
\end{eqnarray}
based on the perturbative assumption that the corrections on the frequency and the damping time are small, i.e. $\delta f \ll f_{\mathrm{GR}}$ and $\delta \tau \ll \delta \tau_{\mathrm{GR}}$, respectively. This kind of test beyond GR has been offically applied to each event analysis, for instance, the loudest event GW250114~\cite{LIGOScientific:2025wao} reported the derivations from the dominant mode $220$ are $\delta f_{220}=0.09^{+0.34}_{-0.22}$ and $\delta \tau_{220} = -0.14^{+0.25}_{-0.23}$.

However, this parametrization method is source-dependent, making it unsuitable for joint constraints across multiple events, and inefficient for data analysis. The further parametrized framework~\cite{Maselli:2019mjd, Carullo:2021dui} seeks a more comprehensive approach to ameliorate the mentioned limitations, in which the frequency and damping time of a mode are expanded directly as polynomial functions of the dimensionless spin $\chi$,
\begin{eqnarray}\label{eq: parameterize formula}
    \omega_K & =&\frac{1}{M_{\mathrm{f}}^{\mathrm{d}}}\sum_{j=0}^{N_{\max }} \chi^j \omega_K^{(j)}\Big(1+\gamma \delta \omega_K^{(j)}\Big)\, ,\nonumber \\
    \tau_K & =&M^{\mathrm{d}}_{\mathrm{f}} \sum_{j=0}^{N_{\max}} \chi^j \tau_K^{(j)}\Big(1+\gamma \delta \tau_K^{(j)}\Big) \, ,
\end{eqnarray}
where $\omega_{K}=2 \pi f_{K}$, $K$ represents the QNMs labels $(l,m,n)$, $\omega^{(j)}_{K}$ and $\tau^{(j)}_{K}$ are the source-independent numerical coefficients based on the spin expansion in GR, these two coefficients have been given in~\cite{Carullo:2021dui}, $M^{\mathrm{d}}_{\mathrm{f}}$ and $\chi$ are the GR-predicted values of the final BH mass and spin in the dector frame, and $N_{\max}$ is the maximum expansion order considered. As a dimensionless coupling constant, $\gamma$ is defined by
\begin{equation}
    \gamma  = \left(\frac{ \ell c^2}{ G M^{\mathrm{s}}_{\rm f}} \right)^p =  \left( \frac{ \ell c^{2} (1+z)}{G M_{\rm f}^{\rm d}} \right)^{p} \, ,
    \label{eq: gamma}
\end{equation}
where $M^{\mathrm{s}}_{\rm f}$ is the final mass in the source frame, $G M^{\mathrm{s}}_{\rm f}/c^2$ is the gravitational radius, and $z$ is the redshift. The parameter $\ell$ sets the characteristic length scale below which corrections beyond GR become relevant, and the exponent $p$ is the power of the scale. All modifications are required to be $\gamma \delta \omega^{(j)}_{K} \ll 1$ and $\gamma \delta\tau^{(j)}_{K} \ll 1$ for all values of $\{K, j\}$ modes considered in the non-extremal spin regime, namely, the perturbative assumption is adopted.

The physical origin of this parameterization method lies in EFTs of GR, whose action in natural units can be described by~\cite{Carullo:2021dui, Silva:2022srr, Cano:2021myl, Cano:2019ore, Cardoso:2018ptl}
\begin{eqnarray}\label{eq: action}
    S_{\mathrm{EFT}} = \frac{1}{16 \pi G} \int \mathrm{d}^4 x \sqrt{-g} \Bigg[
        R + \sum_{n \ge 2} \ell^{2n-2}_{\rm EFT} L_{(2n)}  \Bigg]\, ,
\end{eqnarray}
where $L_{(2n)}$ are the higer-curvature terms containing $2n$ derivitaives,  and $\ell_{\rm EFT} \sim 1/\Lambda_{\mathrm{cutoff}}$ is the energy scale of new physics. Within this framework, corrections associated with higher-curvature terms are encoded in the $\delta \omega^{(j)}_{K}$ and $\delta \omega^{(j)}_{K}$, while the effect of the energy scale $\ell_{\rm EFT}$ and the exponent $2n-2$ are absorbed into $\ell$ and the power scale $p$ both in $\gamma$, respectively. Notably, $\ell_{\rm EFT}$ is not equivalent to the length scale $\ell$ appearing in $\gamma$. The two can differ by a certain factor, since $\ell_{\rm EFT}$ first needs to enter through the corrected Teukolsky equation, and then be translated into the ParSpec parametrization. Besides, a fixed value of $p$ labels a class of modified gravitational theories. For instance, as reported in \cite{Carullo:2021dui}, $p=4$ corresponds to the $L_{(4)}$~($n=2$) operator, and such a quadratic action includes the well-known Einstein-scalar-Gauss-Bonnet~(EsGB) gravity and dynamical Chern-Simon~(dCS) gravity~\cite{Collodel:2019kkx, Alexander:2009tp, Ayzenberg:2014aka}. However, the exponent index $p$ can be rescaled, as long as the alignment relationship is ensured. 
An explicit derivation in EsGB gravity~ \cite{Lyu:2022gdr} and dCS gravity~\cite{Srivastava:2021imr} would clarify the relation between the EFT length scale $\ell_{\rm EFT}$ and the parameter $\ell$ used in ParSpec method.

Current studies~\cite{Maenaut:2024oci,Silva:2022srr,Cano:2021myl} typically perform a joint posterior sampling for the length scale $\ell$ across a set of GW events, and report a $90\%$ or $95\%$ credible upper bound for each fixed value of $p$ that represents a kind of EFT. In these analyses, the ringdown start time and the number of included QNMs are fixed, while regions of parameter space $\gamma<1$ or $\mathrm{Im}(\omega)>0$ are excluded. For instance, the best constraints on the $90\%$ upper bound of $\ell$ jointly posterior distribution for Einstein-scalar-Gauss-Bonnet (EsGB) gravity is $\ell_{\mathrm{EsGB}}< 35\, \mathrm{km}$~\cite{Carullo:2021dui}, for dynamical Chern-Simon (dCS) gravity is $\ell_{\mathrm{dCS}} < 38.7\, \mathrm{km}$~\cite{Silva:2022srr}, for the cubic gravity is $\ell_{\mathrm{c}} \le 38.2\, \mathrm{km}$~\cite{Silva:2022srr}, and for the quartic gravity is $ \ell_{\mathrm{q}} \le 35\, \mathrm{km}$~\cite{Maenaut:2024oci}.

In this work, we dispose of the assumption of a fixed integer $p$, which is proposed in~\cite{Carullo:2021dui}. Formally, we consider 
\begin{equation}
       \gamma  =  \left( \frac{ \tilde{\ell} c^{2} (1+z)}{G M_{\rm f}^{\rm d}} \right)^{\tilde{p}} \, ,
    \label{eq: new gamma}
\end{equation}
where $\tilde{p}$ and $\tilde{\ell}$ replace the $p$ and $\ell$ in Eq. (\ref{eq: gamma}), respectively. Then we perform sampling over both $\tilde{p}$ and $\tilde{\ell}$ simultaneously. If the exponent $\tilde{p}$ is promoted to a free and continuous parameter, correspondingly, one possible interpretation, likewise Eq. (\ref{eq: action}), can be written as
\begin{eqnarray}
    S = \frac{1}{16 \pi G} \int \mathrm{d}^4 x \sqrt{-g} \Bigg[
        R + \int \mathrm{d} \tilde{p} \,  C(\tilde{p}) \, \tilde{\ell}^{\tilde{p}} \mathcal{L}_{(\tilde{p}+2)}  \Bigg] \, .
\end{eqnarray} 
Here, the discrete series of higher-derivative terms is generalized to a continuous representation by replacing the summation with an integral over $\tilde{p}$, from $0$ to an upper limit, and the integrated correction carries the same dimension as the Ricci scalar $R$. $C(\tilde{p})$ is a dimensionless function, by choosing $C(\tilde{p})=\sum_n \alpha_n \delta\left(\tilde{p}-(2n+2)\right)$ and relabeling the summation index such that $n\geq2$, one recovers the standard EFT action. When $\tilde{p}$ is treated as a continuous parameter, it can be interpreted as a fractional scaling index. Correspondingly, $\mathcal{L}_{(\tilde{p}+2)}$, which is a new Lagrangian, may be regarded as an effective rationalized term in the action, incorporating the effects of fractional-order derivatives.

Then, we consider different exclusion thresholds in terms of $\gamma$, ringdown start times, and inclusions of QNMs to assess the impact of the potential parameter space and to test the robustness of the method. Specifically, the events GW250114~\cite{LIGOScientific:2025wao, LIGOScientific:2025rid} and GW231123~\cite{LIGOScientific:2025rsn} are used to perform data analysis; their power spectral sensitivity~(PSD) is estimated following the official procedure in LVK analysis.
The main results are illustrated using GW250114, while GW231123 is incorporated for joint multi-event analysis, which is placed in the Appendix \ref{appendix:GW231123}. Three prior cuts in terms of $\gamma$ are considered: $\gamma<2$, $\gamma<1$, and $\gamma<0.01$, while assigning a uniform prior $\tilde{\ell}\in[0,300]$ and $\tilde{p} \in [0, 20]$ for each case. We analyze two kinds of QNMs inclusion, $220$ and $220+221$. For the $220$-only model, the ringdown start time is varied over $[10.5,20] \, t_{M_{\mathrm{f}}^{\mathrm{d}}}$, whereas for the $220+221$ model it is varied over $[6,15]\,t_{M_{\mathrm{f}}^{\mathrm{d}}}$. In addition, we assign uniform priors $[-0.5,0.5]$ to both $\delta\omega_{K}$ and $\delta\tau_{K}$, and $N_{\rm max}=1$ for all analysis.

We use $\texttt{pyRing}$ to perform Bayesian sampling, in which the standard waveform model is constructed by~\cite{Carullo:2019flw}
\begin{eqnarray}\label{strain}
    h_{+}+\mathrm{i} h_{\times}=\frac{M_{\rm f}^{\rm d}}{D_{\mathrm{L}}} \sum_{l=2}^{\infty} \sum_{m=-l}^{+l} \sum_{n=0}^{\infty}(h_{l m n}^{+}+h_{l m n}^{-})\, ,
\end{eqnarray}
with
\begin{eqnarray}
    h_{l m n}^{+}&=&A_{l m n}^{+} S_{l m n}(\iota, \varphi) \mathrm{e}^{\mathrm{i}\left[\left(t-t_{l m n}\right) \bar{\omega}_{l m n}+\phi_{\ell m n}^{+}\right]}\, ,\\
    h_{l m n}^{-} &=&A_{l m n}^{-} S_{l -m n}(\iota, \varphi) \mathrm{e}^{-\mathrm{i}\left[\left(t-t_{l m n}\right) \bar{\omega}_{l m n}^{\star}-\phi_{l m n}^{-}\right]}\, .
\end{eqnarray}
Here, $\bar{\omega}_{l m n} = 2\pi f_{l m n} + \mathrm{i}/\tau_{l m n}$ is the QNM spectra, and $\star$ represents complex conjugation. For notational simplicity, we omit the subscript $\rm d$ in $M_{\rm f}^{\rm d}$, hereafter $M_{\rm f}$ denotes $M_{\rm f}^{\rm d}$. The $A^{+/-}_{l m n}$ and $\phi^{+/-}_{l m n}$ are the amplitudes and phases of each mode, respectively. The inclination of the BH final spin relative to the observer’s line of sight is denoted by $\iota$, while $\varphi$ corresponds to the azimuthal angle of the line of sight in the BH frame. $S_{l m n}$ are the spin-weighted spheroidal harmonics and $t_{l m n}=t_{0}$ is a reference start time. $D_{\mathrm{L}}$ is the luminosity distance.

When using pyRing for parameter inference, without further assumptions about the underlying theory, one only needs to specify the corresponding QNM frequencies $\bar{\omega}_{lmn}$. In our analysis, we construct these frequencies using the ParSpec framework. ParSpec takes the Kerr QNM frequencies and the remnant parameters as the baseline and encodes deviations from GR into the parameters $\delta\omega_{K}^{(j)}, \delta\tau_{K}^{{j}}$ and $\gamma$. However, both the QNM spectra and the remnant parameters may differ from their GR values in a generic non-GR theory. For example, theory-specific constraints on quadratic or higher-derivative gravity typically use the corresponding modified-theory QNMs~\cite{Chung:2025wbg,Maenaut:2024oci}, in which the deviation parameters from GR $\delta\omega_{K}^{(j)}$ and $\delta\tau_{K}^{j}$ are fixed. Our choice is motivated by two considerations. One is that allowing $\tilde{p}$ to vary freely prevents a direct mapping to any particular gravity theory. The other is based on the perturbative assumption, in which higher-curvature corrections only weakly shift the Kerr QNMs in the non-extremal spin regime. The remnant spin is inferred to be $\chi_{\mathrm{f}}=0.68^{+0.01}_{-0.01}$ for GW250114~\cite{LIGOScientific:2025rid, LIGOScientific:2025wao} and $\chi_{\mathrm{f}}=0.82^{+0.06}_{-0.11}$ for GW231123~\cite{LIGOScientific:2025rsn} (\textsc{NRSur7dq4}), these make the use of Kerr QNMs available.

Methodologically, to alleviate the degeneracy with the luminosity distance, we restrict $D_{\mathrm{L}}$ to the $95\%$ credible interval inferred from the \textsc{NRSur7dq4} posterior distribution~\cite{Varma:2019csw}. We also adopt the prior distribution of the remnant mass $M_{\mathrm{f}}$ predicted by inspiral-merger, in order to reduce the additional degeneracy. These two settings in sampling act as informative priors in the data analysis~\cite{Wang:2026rev,Chen:2025wfi}.

\section{Results and discussions}\label{sec: Results and Discussions}

\subsection{Informative prior}
\begin{figure}
    \centering
    \includegraphics[width=0.9\linewidth]{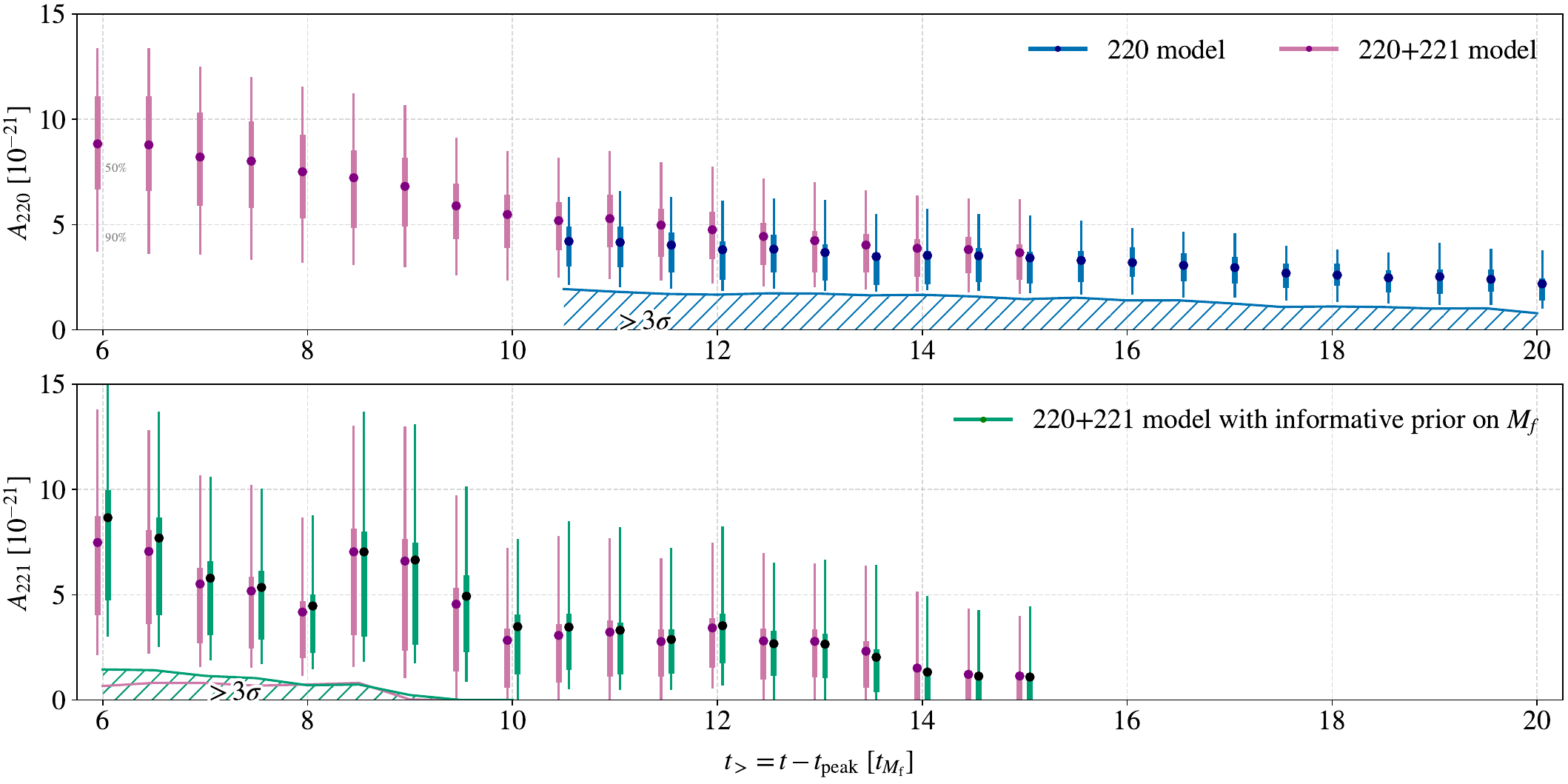}
    \caption{The inferred results on QNMs amplitudes with GW250114, provided by $\texttt{pyRing}$ package. The $220$-only model is represented by blue markers, with the ringdown start time scanned over $[10.5,20]t_{M_{\mathrm{f}}}$. The $220+221$ model without informative priors is shown by pink markers, with the start time varied within $[6,15]t_{M_{\mathrm{f}}}$. The corresponding $220+221$ analysis with informative priors is indicated by green markers, using the same start-time range as the case without informative priors. In all three cases, the start time is scanned with an interval of $0.5\,t_{M_{\mathrm{f}}}$. The black central points indicate the median of the posterior samples, while the thick and thin error bars denote the $50\%$ and $90\%$ credible intervals, respectively. The shaded region corresponds to the parameter space excluded at the $3\sigma$ confidence level.}
    \label{fig:inferred QNMs amplitude results.}
\end{figure}

We first validate the role of the informative priors by examining their impact on the inferred QNM strain amplitudes with GW250114, especially for the fundamental mode and the first overtone. Fig. \ref{fig:inferred QNMs amplitude results.} shows the posterior distributions of the QNMs strain amplitudes. The results of the $220$-only model are represented by blue markers, with the ringdown start time scanned over $[10.5,20] \, t_{M_{\mathrm{f}}}$. The results of the $220+221$ model without informative priors are shown by pink markers, with the start time varied within $[6,15]\,t_{M_{\mathrm{f}}}$. The corresponding $220+221$ analyses with informative priors are indicated by green markers, using the same start-time range as the case without informative priors. In all three cases, the start time is scanned with an interval of $0.5\,t_{M_{\mathrm{f}}}$. The black central points indicate the median of the posterior samples, while the thick and thin error bars denote the $50\%$ and $90\%$ credible intervals, respectively. The shaded region corresponds to the parameter space excluded at the $3\sigma$ confidence level. 

We find that the inferred strain amplitudes of the fundamental mode and first overtone are qualitatively consistent with the official analysis~\cite{LIGOScientific:2025rid}, which are produced by the $\texttt{ringdown}$ package. After imposing an informative prior on $M_{\mathrm{f}}$, the constraints on the first overtone amplitude become noticeably tighter at early ringdown times. This improvement gradually weakens at later times, as the contribution of the $221$ mode to the signal decreases. Eventually, the overtone signal becomes too weak, causing the difference between analyses with and without informative priors to disappear. The origin of this effect can be understood from the \texttt{pyRing} waveform model, in which the strain amplitude is determined by the combination $\frac{M_{\rm f}}{D_{\mathrm{L}}} A_{l m n}^{\pm}$. As a result, tightening the parameter space of $M_\mathrm{f}$  and $D_{\mathrm{L}}$ suppresses parameter degeneracies and improves the constraints on the strain amplitudes. We adopt the same strategy for sampling $\gamma$, which is determined by $M_{\mathrm{f}}$, $\tilde{\ell}$, and $\tilde{p}$. Imposing the informative prior on $M_{\mathrm{f}}$ reduces its degeneracy with $\tilde{\ell}$ and is expected to yield a better-constrained range of $\tilde{\ell}$, $\tilde{p}$, and $\gamma$ in the subsequent analysis.

\subsection{Parameter space}
\begin{figure}
    \centering
    \includegraphics[width=0.8\linewidth]{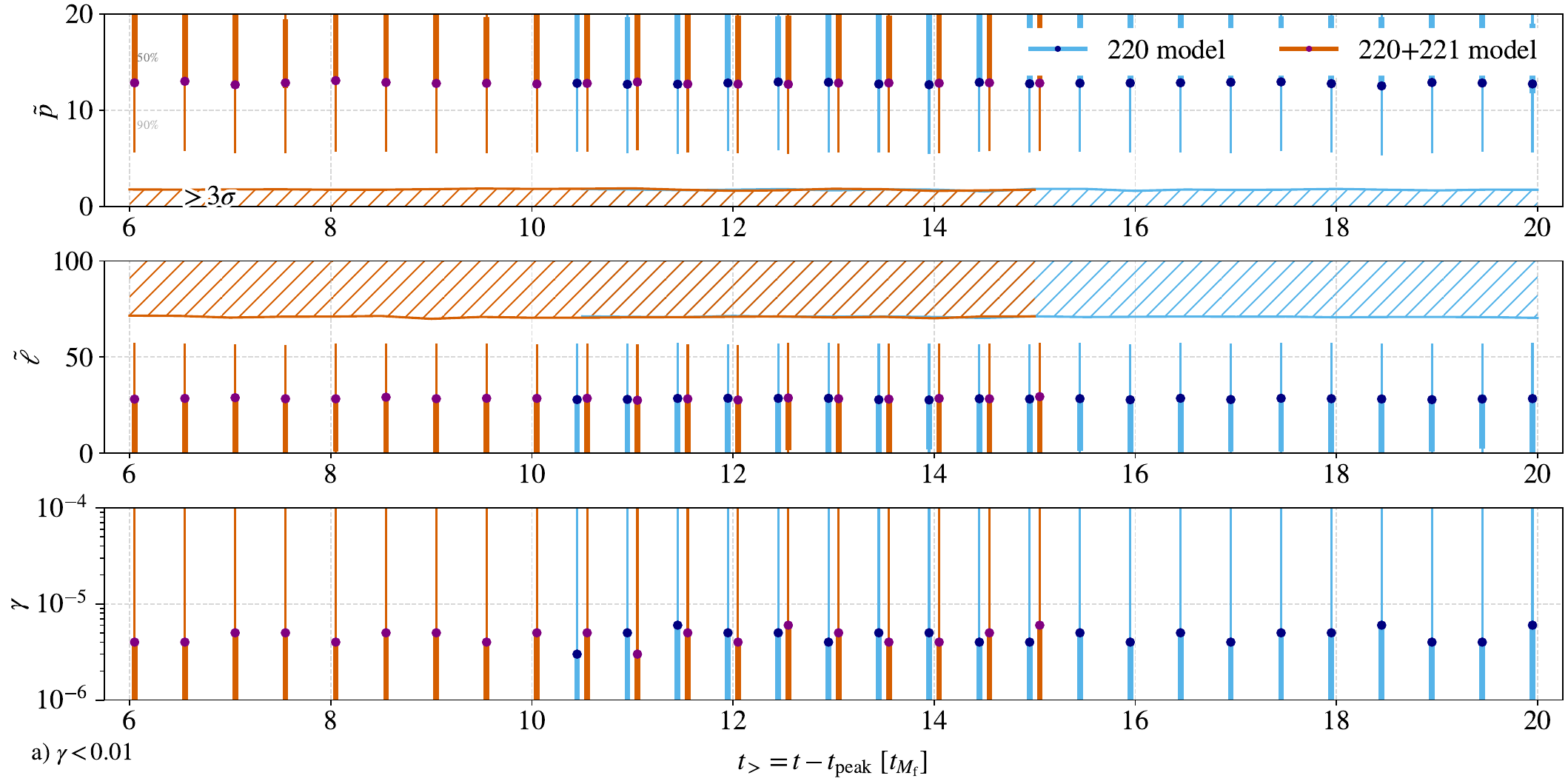}
    \includegraphics[width=0.8\linewidth]{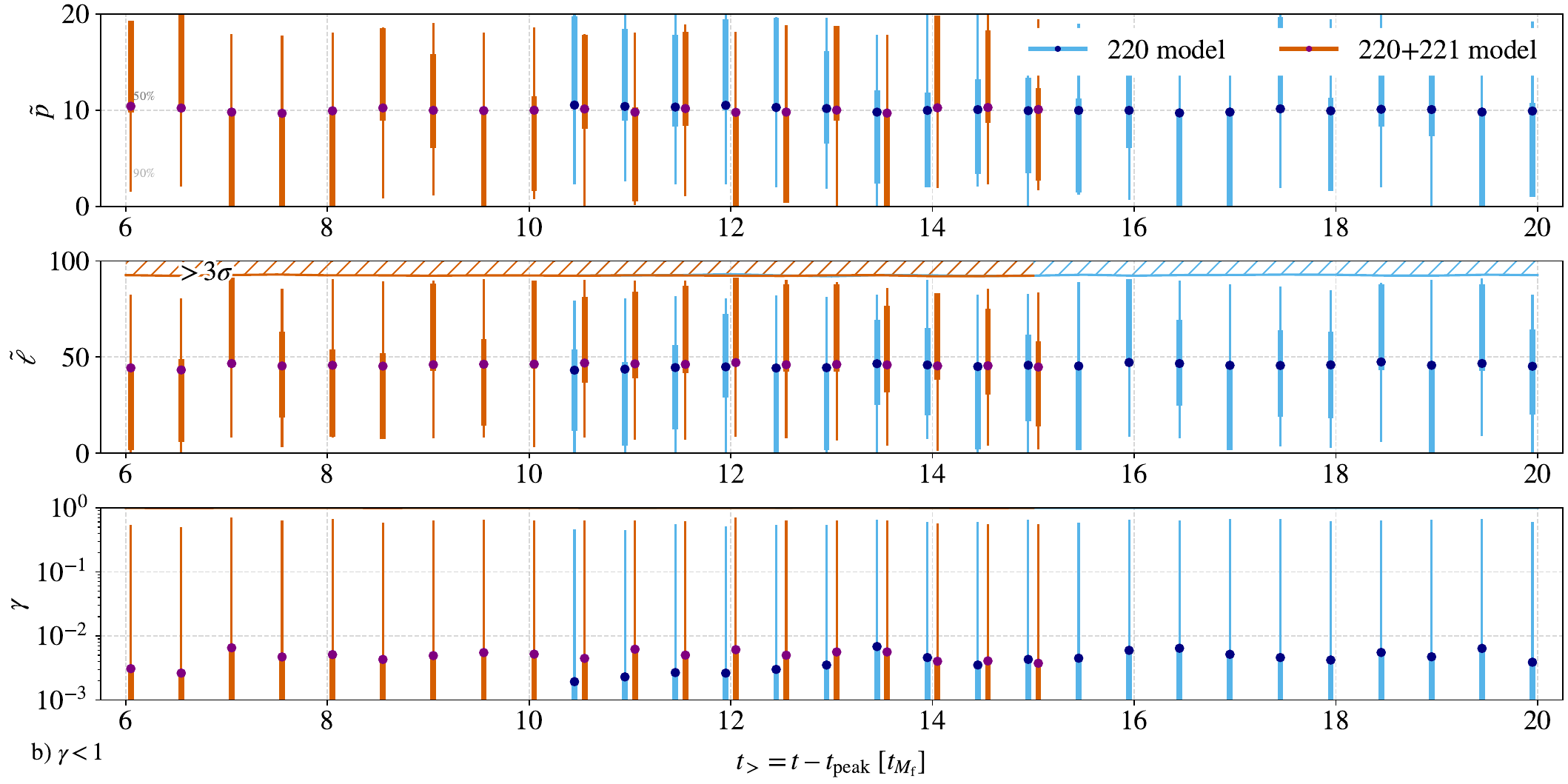}
    \includegraphics[width=0.8\linewidth]{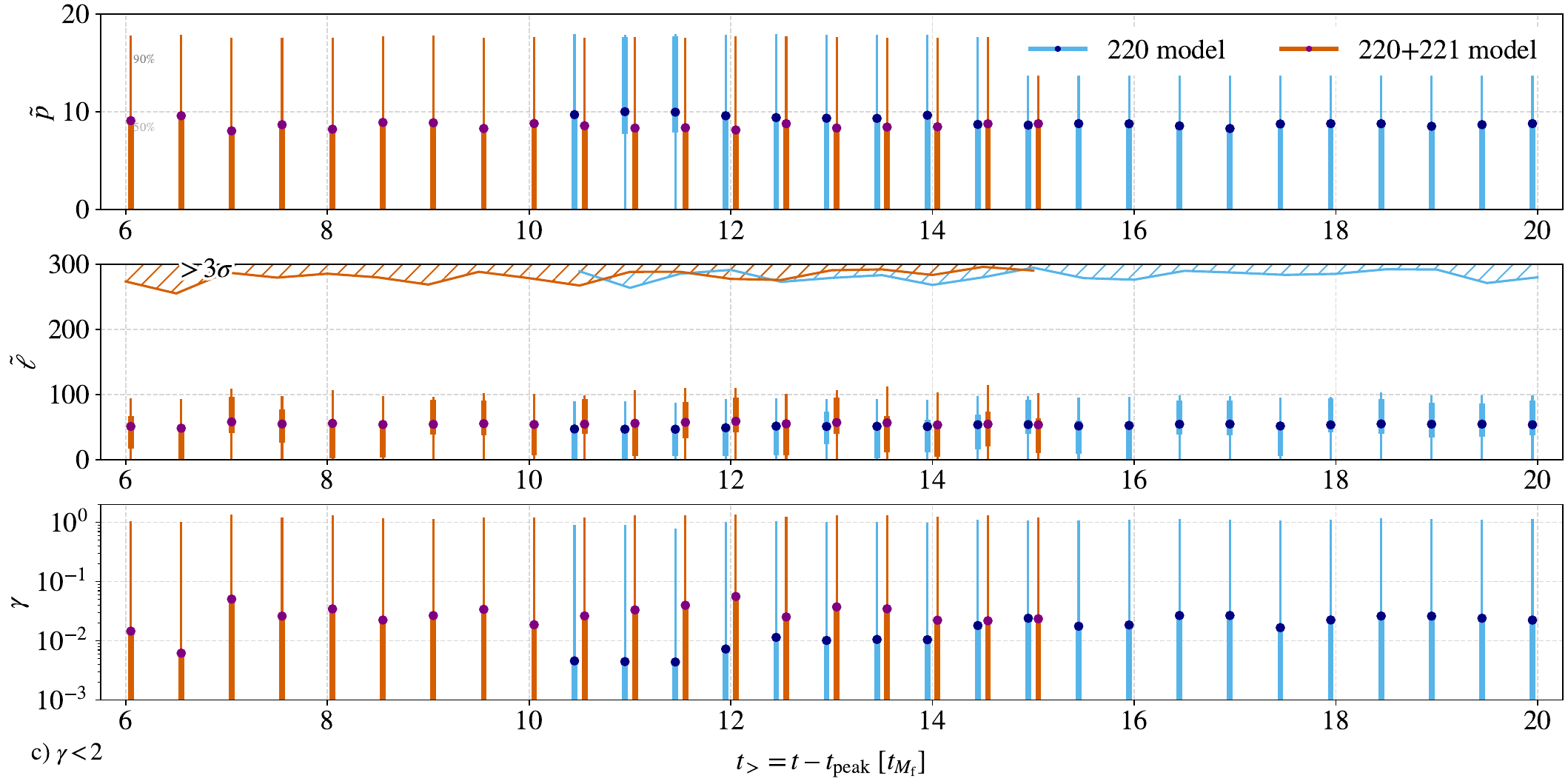}
    \caption{The posterior distributions of $\tilde{p}$, $\tilde{\ell}$, and $\gamma$ with GW250114 under $\tilde{p} \in [0,20]$, provided by $\texttt{pyRing}$ package. The three pannels correspond to the three prior cuts for $\gamma$ $a) \, \gamma<0.01$, $b) \, \gamma<1$, and $c) \, \gamma<2$. In each case, the $220$-only results are shown in light blue, and the $220+221$ results are shown in orange. In all three cases, the posterior distributions are shown as functions of the ringdown start time, and it is scanned with an interval of $0.5\,t_{M_{\mathrm{f}}}$. The black central points indicate the median of the posterior samples, while the thick and thin error bars denote the $50\%$ and $90\%$ credible intervals, respectively. The shaded region corresponds to the parameter space excluded at the $3\sigma$ confidence level.}
    \label{fig:main results}
\end{figure}

Fig. \ref{fig:main results} presents the GW250114 constraints obtained for  $\tilde{p} \in [0,20]$. The three panels correspond to the three prior cuts in terms of $\gamma$, $a) \, \gamma<0.01$, $b) \, \gamma<1$, and $c) \, \gamma<2$, marked at the bottom left of each panel. In each case, the $220$-only results are shown in light blue, and the $220+221$ results are shown in orange; the posterior distributions of $\tilde{p}$, $\tilde{\ell}$, and $\gamma$ are shown as functions of the ringdown start time. The perturbative consistency condition $\gamma \delta \omega^{(j)} \ll 1$ and $\gamma \delta\tau^{(j)}$ are explicitly checked throughout all analyses. The remaining plotting conventions are the same as those adopted in Fig. \ref{fig:inferred QNMs amplitude results.}. The analyses for the posterior distributions are summarised below.

\begin{figure}
    \centering
    \includegraphics[width=0.9\linewidth]{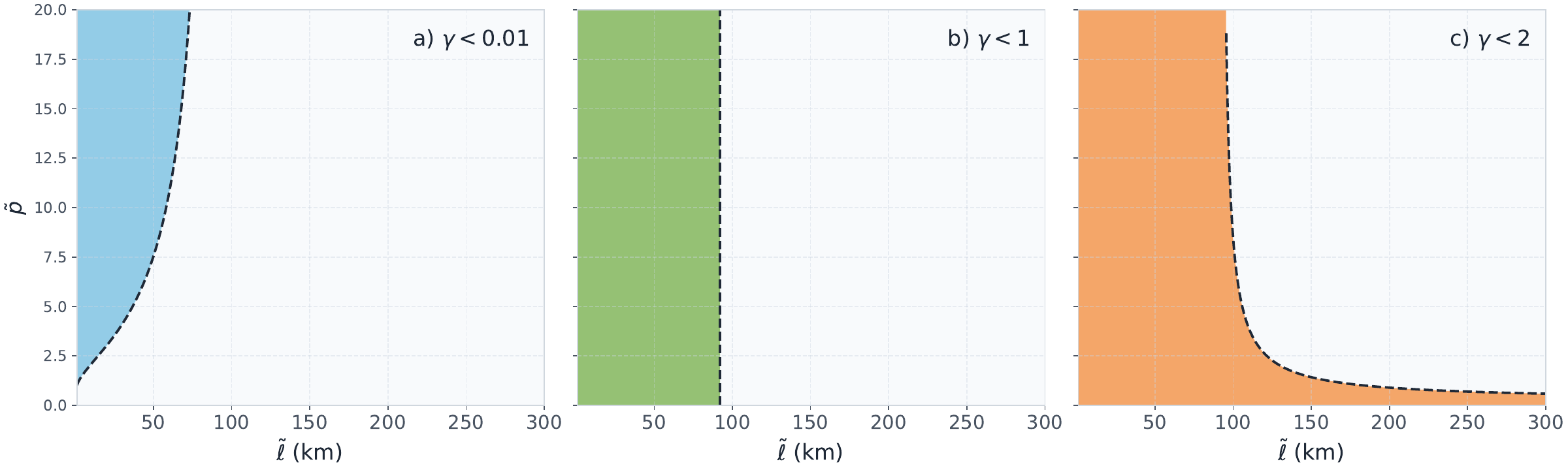}
    \caption{Allowed regions in the $(\tilde{\ell},\tilde{p})$ parameter space under different exclusion prescriptions on $\gamma$: a) $\gamma<0.01$, b) $\gamma<1$, and c) $\gamma<2$. The colored regions indicate the parameter space satisfying the corresponding condition. The dashed black curves represent the associated boundaries, where we use $M_{\mathrm{f}}=68,z=0.09$ and $\tilde{\ell} \in [0, 300], p \in [0, 20]$.}
    \label{fig:parameter space}
\end{figure}

An apparent feature in panel a)\,$\gamma<0.01$ of Fig. \ref{fig:main results} is that the small-$\tilde{p}$ region is excluded at $3\sigma$ significance. This characteristic should not be interpreted as evidence against lower-order higher-curvature gravities. Instead, it arises from the restrictions of the parameter space induced by the prior conditions of $\gamma$. Fig. \ref{fig:parameter space} explicitly shows the allowed regions in the $(\tilde{\ell},\tilde{p})$ plane under different prior choices of $\gamma$, where we use $M_{\rm f}=68,z=0.09$ and $\tilde{\ell} \in [0, 300], \tilde{p} \in [0, 20]$. Different prior prescriptions generate qualitatively different geometries of the allowed parameter space. For the condition $\gamma<1$, the allowed parameter space remains uniform in both $\tilde{p}$ and $\tilde{\ell}$, sampling over $\tilde{p}$ introduces little bias into the inferred $\tilde{\ell}$ constraints. However, the stringent condition $\gamma<0.01$ strongly suppresses the low-$\tilde{p}$ region; the distorted parameter space leads to the exclusion of the small-$\tilde{p}$ region in the posterior distribution, as shown in panel a) of Fig. \ref{fig:main results}. For the weaker condition $\gamma<2$, larger values of $\tilde{\ell}$ are only permitted for sufficiently small $\tilde{p}$. Thus, the posterior distribution of $\tilde{p}$ shown in panel c) of Fig. \ref{fig:main results} is shifted toward smaller values, whereas the constraint on $\tilde{\ell}$ mainly excludes only the larger-$\tilde{\ell}$ region, compared to the results in panels a) and b). In fact, once the upper bound on $\gamma$ is chosen away from unity, the allowed $(\tilde{\ell},\tilde{p})$ parameter space becomes distorted. The allowed range of $\tilde{\ell}$ will depend explicitly on $\tilde{p}$, and vice versa. From this perspective, the condition $\gamma<1$ is not simply one possible prior cut among others, but the most natural consistency condition for avoiding an artificial correlation. Furthermore, to verify whether the inferred $\tilde{\ell}$ posterior is affected by the prior range of $\tilde{p}$, we repeat the above sampling under $\tilde{p} \in [0,10]$. Those results are shown in Appendix \ref{apx:the results of the prior choice}, indicating that the prior upper bound of $\tilde{p}$ will affect the sampling results of $\tilde{\ell}$, when the $\gamma$ condition deviates from $\gamma<1$.

Based on the discussion on the parameter space, the most representative posterior distributions are derived from the $\gamma<1$ case. In this case, the inferred $\tilde{p}$ in Fig. \ref{fig:main results} is broad and nearly prior-dominated. This behaviour indicates that the current ringdown data have limited sensitivity to this scaling index, and are unable to distinguish different EFT scaling. Posterior results of three parameters also show that different values of $\tilde{p}$ can generate similar $\gamma$ shifts through compensating changes in $\tilde{\ell}$. From Eq. (\ref{eq: gamma}), if the coupling condition is rewritten as
\begin{eqnarray}
    \ln\gamma= \tilde{p} \ln\Bigg(\frac{\tilde{\ell}c^2 (1+z)}{G M_{\rm f}}\Bigg) < 0 \, ,
\end{eqnarray}
the degenercy between $\tilde{p}$ and $\tilde{\ell}$ and the different constraining power on the two parameters become more transparent in this form. Moreover, a notable feature is the emergence of a plateau region for the fundamental length scale $\tilde{\ell}$, about $\tilde{\ell} \sim 0-90\, \mathrm{km}$. Values outside of this region are excluded at the $3\sigma$ significance level, which is consistent with the analysis of the $(\tilde{\ell}, \tilde{p})$ parameter space. For the parameter $\gamma$, the posterior distribution consistently favors the small-coupling regime, indicating no evidence for deviations from the QNM spectra predicted by GR. Additionally, in the overlapping time range of the $220+221$ model and the $220$-only model, the mean of $\gamma$ posterior distributions obtained from the two analyses gradually converge. 

\subsection{Starting time and mode content}
\begin{figure}
    \centering
    \includegraphics[width=0.45\linewidth]{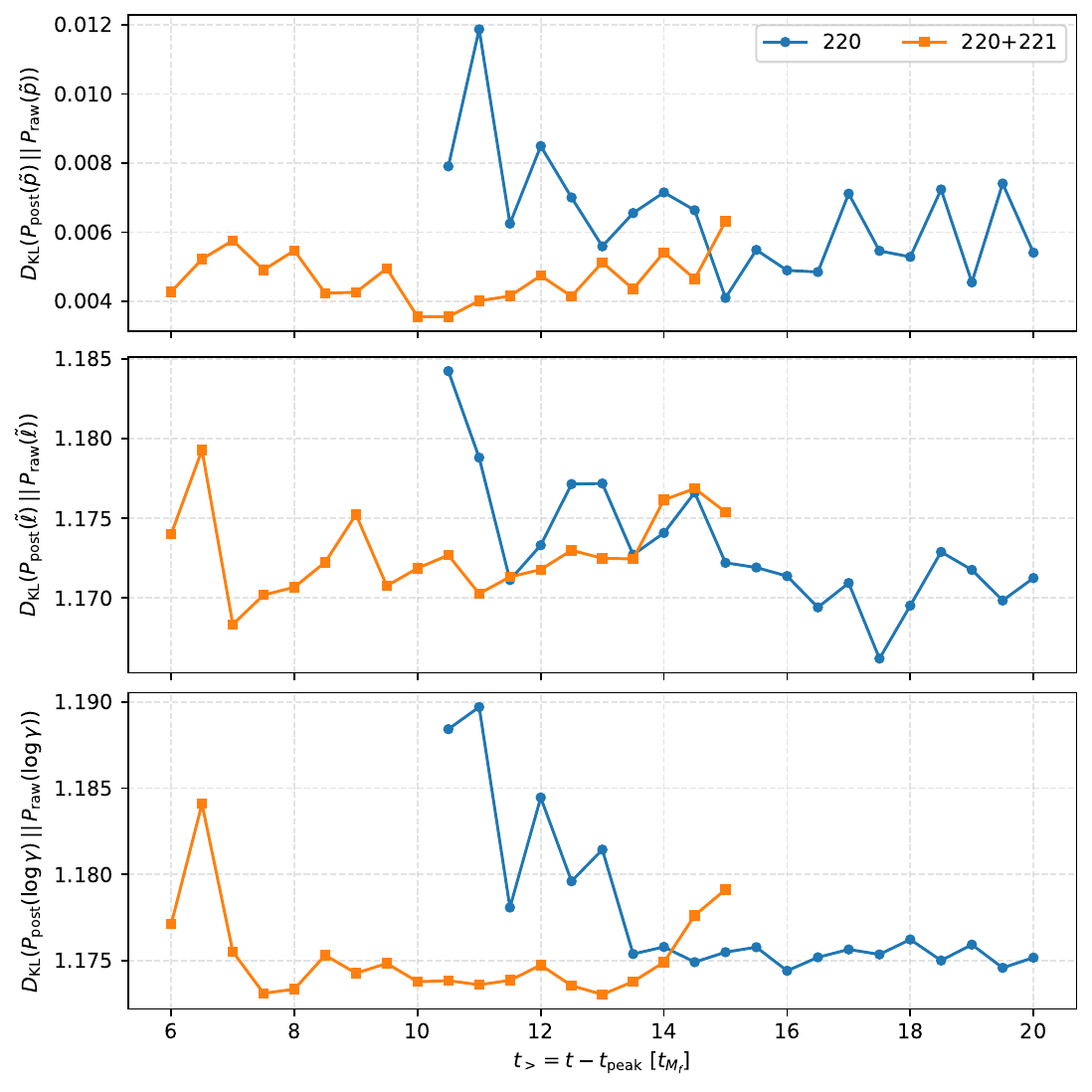}
    \includegraphics[width=0.45\linewidth]{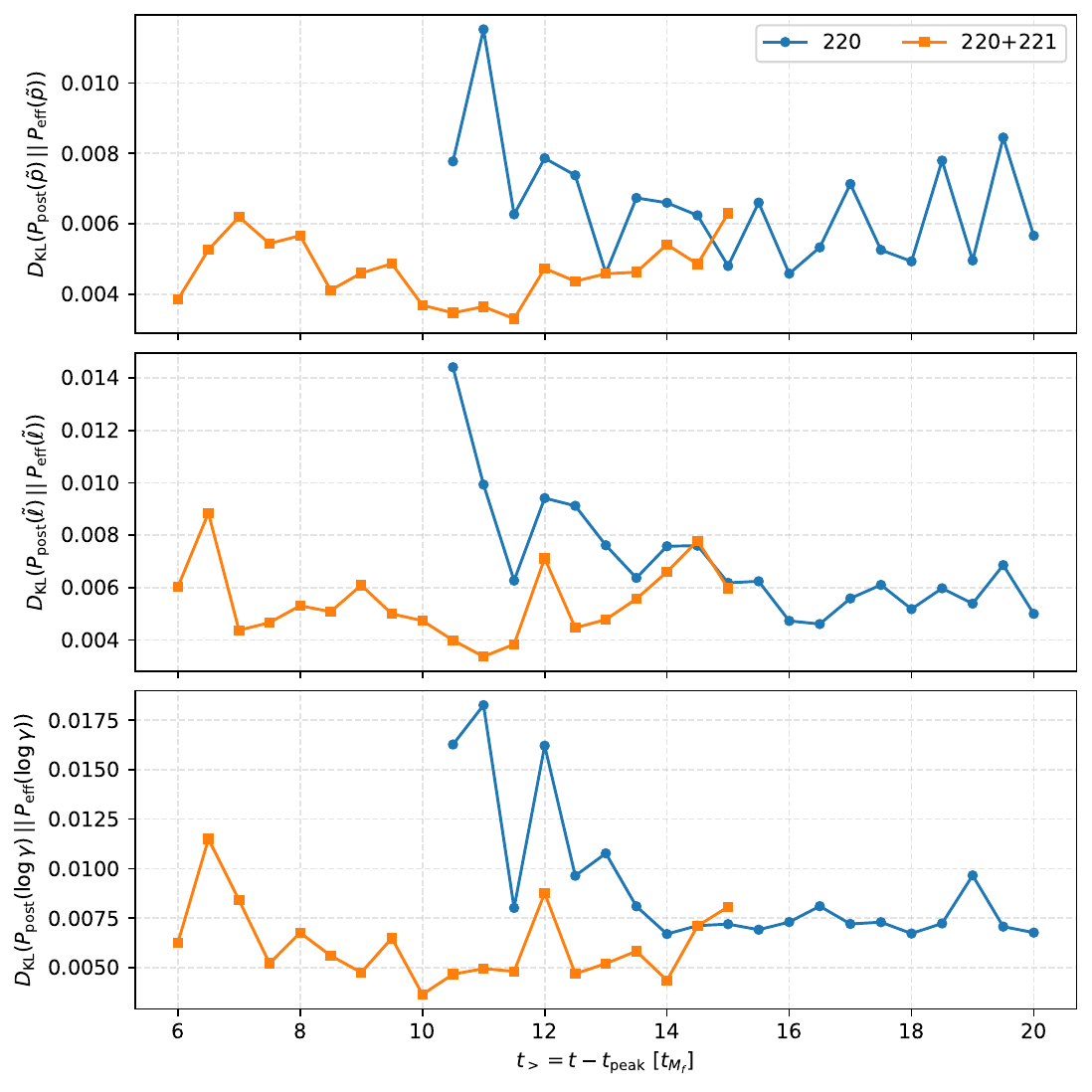}
    \caption{KL divergences for the posterior distributions of $\tilde{p}$, $\tilde{\ell}$, and $\log\gamma$ as functions of the ringdown start time $t_{\geq}=t-t_{\rm peak}$. The blue and orange curves correspond to the $220$ and $220+221$ models, respectively. The left panel shows the KL divergences computed with respect to the raw priors, namely $[0,20]$ for $\tilde{p}$, $[0,300]\, {\rm km}$ for $\tilde{\ell}$, and the induced raw prior $\log\gamma$. The right panel shows the corresponding KL divergences with respect to the effective priors $P_{\rm eff}$, obtained by applying the condition $\gamma<1$ to samples drawn from the raw priors.}
    \label{fig:KL divergence}
\end{figure}

To quantify the information gain from the ringdown data for different choices of starting time and QNM mode content, we compute the Kullback-Leibler (KL) divergence, which is defined as follows
\begin{eqnarray}\label{KL_divergence}
    D_{\rm KL}(t_{0}^{i}; P_{\rm post}(\tilde{\theta})||P_{\rm prior}(\tilde{\theta}))=
    \int \mathrm{d} \tilde{\theta} \, P_{\rm post}(\tilde{\theta}| \mathrm{d},t_{0}^{i})
    \ln\frac{P_{\rm post}( \tilde{\theta} | \mathrm{d},t_{0}^{i})}{P_{\rm prior}(\tilde{\theta})} \, ,
\end{eqnarray} 
where $\tilde{\theta}$ denotes the parameter to be compared, $t_{0}^{\rm i}$ represents the each start time, $\rm d$ means data, $P_{\rm post}$ is the posterior distribution, and $P_{\rm prior}$ is the prior distribution. We compute the KL divergence for $\tilde{\theta}={\tilde{p},\tilde{\ell},\log\gamma}$. Two kinds of reference priors are considered. The first is the raw prior, namely $\tilde{p}\in[0,20]$, $\tilde{\ell}\in[0,300]\, {\rm km}$, and the induced raw prior of $\log\gamma$. The second is the effective prior $P_{\rm eff}$, obtained by drawing samples from the raw priors, computing $\gamma=\left[\frac{\tilde{\ell}c^2(1+z)}{G M_{\rm f}}\right]^{\tilde{p}}$ and retaining only samples satisfying $\gamma<1$. This effective prior accounts for the prior-volume reduction caused by the prior condition in terms of $\gamma$. Fig. \ref{fig:KL divergence} presents the KL divergences computed with respect to two choices of reference prior. The left panel shows the results relative to the raw prior, and the right panel shows those relative to the effective prior. The blue and orange curves denote the \(220\) and \(220+221\) models, respectively. The KL divergence for $\tilde{p}$ remains small in both the raw-prior and effective-prior comparisons, confirming that the current data provide little information on the EFT scaling index. By contrast, the raw-prior case gives $\mathcal{O}(1)$ KL divergences for $\tilde{\ell}$ and $\log\gamma$. These values are substantially reduced to the $\mathcal{O}(10^{-2})$ level once the effective prior is adopted. This reduction indicates that the primary information gain in $\tilde{\ell}$ and $\log\gamma$ is mainly caused by the prior-volume reduction induced by the $\gamma<1$ condition, rather than by the ringdown data itself.

Although the KL divergences remain small in magnitude, a noteworthy feature can be observed in the overlapping time range of the two analyses. The $220$-only model yields larger KL divergences than the $220+221$ model, while this difference gradually diminishes as the ringdown start time is shifted later. A similar tendency can also be observed in the posterior distribution of \(\gamma\) in panel b) of Fig. \ref{fig:main results}, where the $220$-only result starts from smaller values and gradually converges to the $220+221$ result at later times. A possible interpretation for these two behaviors is that the $221$ mode is no longer statistically supported in the overlap time range. At early start times of this data segment, the Signal-to-Noise Ratio (SNR) is relatively high, so the \(220\)-only model imposes a stronger constraint on $\gamma$, leading to a relatively larger KL divergence. By contrast, including the $221$ mode introduces additional degrees of freedom into the model. If the data do not contain a statistically significant $221$ contribution, these extra degrees of freedom mainly broaden the inference. As a result, the $\gamma$ posterior can be allowed relatively larger, and the KL divergence can become smaller. As the start time shifts later, the remaining ringdown SNR decreases, and the constraining power on the $\gamma$ weakens. Consequently, the KL divergence obtained from the $220$-only analysis decreases, and both the $\gamma$ posterior and the KL divergence gradually converge between the two models.

\begin{figure}
    \centering
     \includegraphics[width=0.9\linewidth]{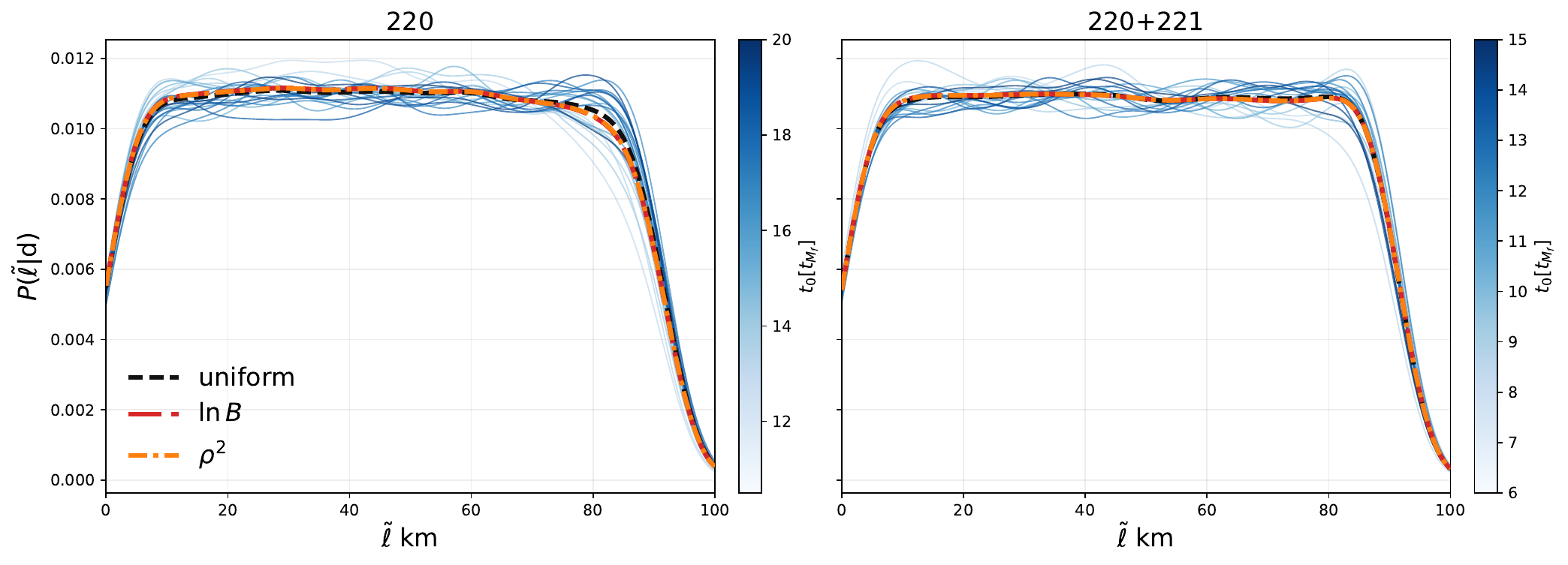}
    \caption{
    One-dimensional posterior distributions of the characteristic length scale $\tilde{\ell}$ with $\tilde{p}<20$ for the $220$ and $220+221$ models under the representative condition $\gamma<1$. The thin blue curves show the posteriors obtained at individual ringdown start times $t_0$, with the colorbar indicating the value of $t_0[t_{M_{\rm f}}]$. The thick curves denote the posteriors marginalized over $t_0$ using three weighting prescriptions: the black dashed curve corresponds to uniform weighting over the selected start-time range, the red dash-dotted curve corresponds to $\ln B$-based weighting, where $\ln B = \ln (Z_{\rm signal} / Z_{\rm noise})$ and the orange dash-dotted curve corresponds to $\rho^2$-based weighting, where $\rho$ is the estimated ringdown SNR.}
    \label{fig:marginalized ell}
\end{figure}

To further assess the impact of the ringdown start time and mode content on the constraint of $\tilde{\ell}$, we overplot the one-dimensional posterior distribution $P(\tilde{\ell}|\mathrm{d},t_{0}^{i})$ obtained at each start time for each mode content.  Then we construct posterior distributions marginalized over the ringdown start time $t_0$, using
\begin{eqnarray}
    P(\tilde{\ell}|\mathrm{d})= \sum_{i} w_{i} P(\tilde{\ell}|\mathrm{d},t_0^i) \, ,
\end{eqnarray}
where $w_{i}$ is the weight. This marginalization is subtle because different start times imply different data segments of the same signal. We therefore adopt three weighting prescriptions: uniform weighting over the selected start-time range, $\ln B$-based weighting, where $\ln B = \ln (Z_{\rm signal} / Z_{\rm noise})$, and $\rho^2$ weighting based on the SNR $\rho$. As shown in Fig. \ref{fig:marginalized ell}, the thin blue curves show the one-dimensional posteriors, with the colorbar indicating the value of $t_0$. In each panel, the black dashed, red dash-dotted, and orange dash-dotted curves correspond to uniform, $\ln B$-based, and $\rho^2$-based weighting, respectively. The apparent left boundary features are introduced by the Gaussian kernel density estimation (KDE) procedure used to smooth the posterior samples. 

Then we compute the $90\%$ credible upper bounds $\tilde{\ell}_{90}$, which is defined by $\int^{\tilde{\ell}_{90}}_{0} \mathrm{d} \tilde{\ell} \, P(\tilde{\ell}|\mathrm{d})=0.9$, in the overlapping time range of the two analyses, and summarize the corresponding differences $\Delta\tilde{\ell}_{90}$ in Tab. \ref{tab:ell90_same_t0_comparison}. The results show that the values of $\tilde{\ell}_{90}$ remain nearly unchanged across different start times within each mode content, staying around $\sim 80 \, {\rm km}$ with no significant variation. At the same start time, the $220$-only and $220+221$ models show small differences in $\tilde{\ell}_{90}$ at earlier times, but these differences gradually decrease as the start time is shifted later. This weak dependence on both $t_0$ and QNMs content should be interpreted with discretion. On the one hand, the KL divergence based on the effective prior shows the limited constraining power on $\tilde{\ell}$, so that the $\tilde{\ell}$ posterior remains largely dominated by the parameter space allowed by the imposed $\gamma<1$ condition. On the other hand, the similarity of the values also indicates that the framework gives stable measurement under different choices of start time and QNMs content. Thus, we interpret the results as a stable but weak constraint on $\tilde{\ell}$, and the stability currently arises from the parameter-space restriction rather than from ringdown data. Together with the behavior of the KL divergence in the overlapping time range, we conclude that, under the limited SNR of the current data, a reliable single-mode model can provide a more conservative and stable constraint within the ParSpec framework at the ringdown start time, where the SNR of the single-mode remains relatively high.

\begin{table}[htbp]
\centering
\small
\setlength{\tabcolsep}{8pt}
\renewcommand{\arraystretch}{1.12}
\begin{tabular}{c c c c}
\hline
\hline
$t_0 \, [t_{M_{\rm f}}]$
& $\tilde{\ell}_{90}^{220}\,[{\rm km}]$ 
& $\tilde{\ell}_{90}^{220+221}\,[{\rm km}]$ 
& $\Delta\tilde{\ell}_{90}\,[{\rm km}]$ \\
\hline
10.5 & 79.511 & 82.861 & -3.350 \\
11.0 & 80.744 & 83.046 & -2.302 \\
11.5 & 81.609 & 83.149 & -1.540 \\
12.0 & 80.826 & 83.875 & -3.049 \\
12.5 & 81.583 & 82.838 & -1.255 \\
13.0 & 81.680 & 82.856 & -1.176 \\
13.5 & 82.576 & 82.721 & -0.146 \\
14.0 & 82.922 & 81.945 &  0.977 \\
14.5 & 82.371 & 82.046 &  0.325 \\
15.0 & 82.591 & 81.961 &  0.629 \\
\hline
\hline
\end{tabular}
\caption{
Comparison of the $90\%$ credible upper bounds $\tilde{\ell}_{90}$ obtained from the $220$ and $220+221$ models at the same ringdown start times in the overlapping time range. The difference is defined as $\Delta\tilde{\ell}_{90}=\tilde{\ell}_{90}^{220}-\tilde{\ell}_{90}^{220+221}$. The small differences indicate that the inferred upper bound on $\tilde{\ell}$ is weakly sensitive to the inclusion of the $221$ overtone within the overlapping start-time range.}
\label{tab:ell90_same_t0_comparison}
\end{table}

\subsection{Degeneracy between \texorpdfstring{$\tilde{\ell}$ and $\tilde{p}$}{ell and p}}

The central challenge of this approach is the degeneracy between $\tilde{\ell}$ and $\tilde{p}$. Measuring additional QNMs or jointly analyzing multiple events offers promising ways to break this degeneracy. At the SNR level of the currently available events, accurately resolving several QNMs remains challenging. Therefore, a joint analysis of multiple events with different remnant masses provides a more practical way to reduce the $\ell-p$ degeneracy. In the following analysis, we include GW231123 to illustrate how joint constraints can be obtained from multiple events. More details about the data processing of GW231123 can be seen in Appendix \ref{appendix:GW231123}.

\begin{figure}
    \centering
    \includegraphics[width=0.9\linewidth]{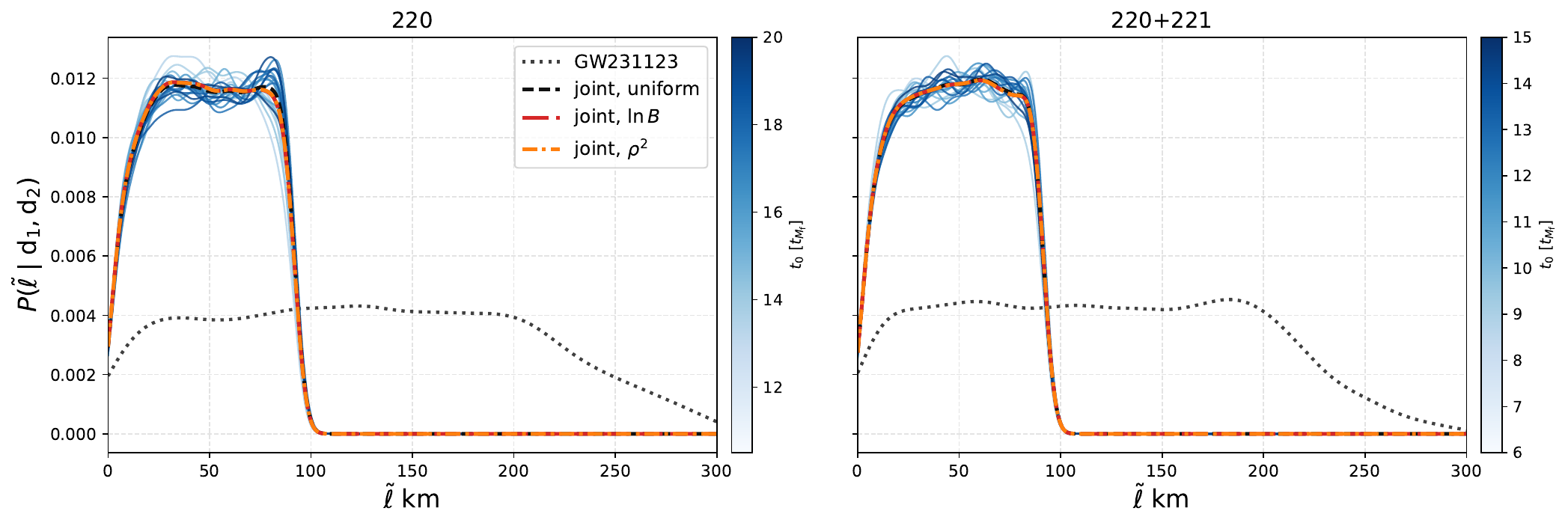}
    \caption{Joint posterior distributions of the characteristic length scale $\tilde{\ell}$ obtained by combining GW231123 with GW250114 under the representative condition $\gamma<1$. The left and right panels correspond to the $220$ and $220+221$ models, respectively. The thin blue curves show the joint posteriors obtained at individual ringdown start times $t_0$, with the colorbar indicating $t_0[t_{M_{\rm f}}]$. The gray dotted curve denotes the posterior obtained from GW231123. The thick curves show the joint posteriors marginalized over $t_0$ under three weighting prescriptions: black dashed for uniform weighting, red dash-dotted for $\ln B$-based weighting, and orange dash-dotted for \(\rho^2\)-based weighting, where \(\rho\) is the estimated ringdown SNR.}
    \label{fig:joint_event_posterior}
\end{figure}

We first examine the effect of adding GW231123 through a joint constraint on the characteristic length scale $\tilde{\ell}$, as shown in Fig. \ref{fig:joint_event_posterior}. For the joint constraint, we treat $\tilde{\ell}$ as a common source-independent parameter shared by GW250114 and GW231123, while the event-specific parameters are marginalized independently for each event. The joint posterior $P_{\rm joint}$ is then constructed from the product of the independent marginalized likelihoods,
\begin{eqnarray}\label{Joint}
    P_{\rm joint}(\tilde{\ell}|\mathrm{d}_1,\mathrm{d}_2,t_{0}^{i}) \propto \pi(\tilde{\ell})\, \mathcal{L}_1(\mathrm{d}_1|\tilde{\ell},t_{0}^{i} )\, \mathcal{L}_2(\mathrm{d}_2|\tilde{\ell},t_{0}^{i}) \, ,
\end{eqnarray}
where $\pi(\tilde{\ell})$ is the common prior and $\mathcal{L}_i(\mathrm{d}_i|\tilde{\ell},t_{0}^{i})$ denotes the likelihood marginalized over the event-specific parameters. The left and right panels correspond to the $220$ and $220+221$ models, respectively. The gray dotted curve denotes the GW231123-only posterior, while the thin blue curves show the joint posteriors with GW250114 at its ringdown start times indicated by the colormap. The other colored curves follow the convention used in Fig. \ref{fig:marginalized ell}. The GW231123-only posterior is relatively broad and contains a pronounced large-$\tilde{\ell}$ tail. After combining it with GW250114, this large-$\tilde{\ell}$ tail is strongly suppressed, indicating that the large-$\tilde{\ell}$ region is not coherently supported by the two events. However, the resulting joint posterior largely overlaps with the GW250114 posterior, showing that the combined constraint is primarily driven by GW250114.

\begin{figure}
    \centering
    \includegraphics[width=0.48\linewidth]{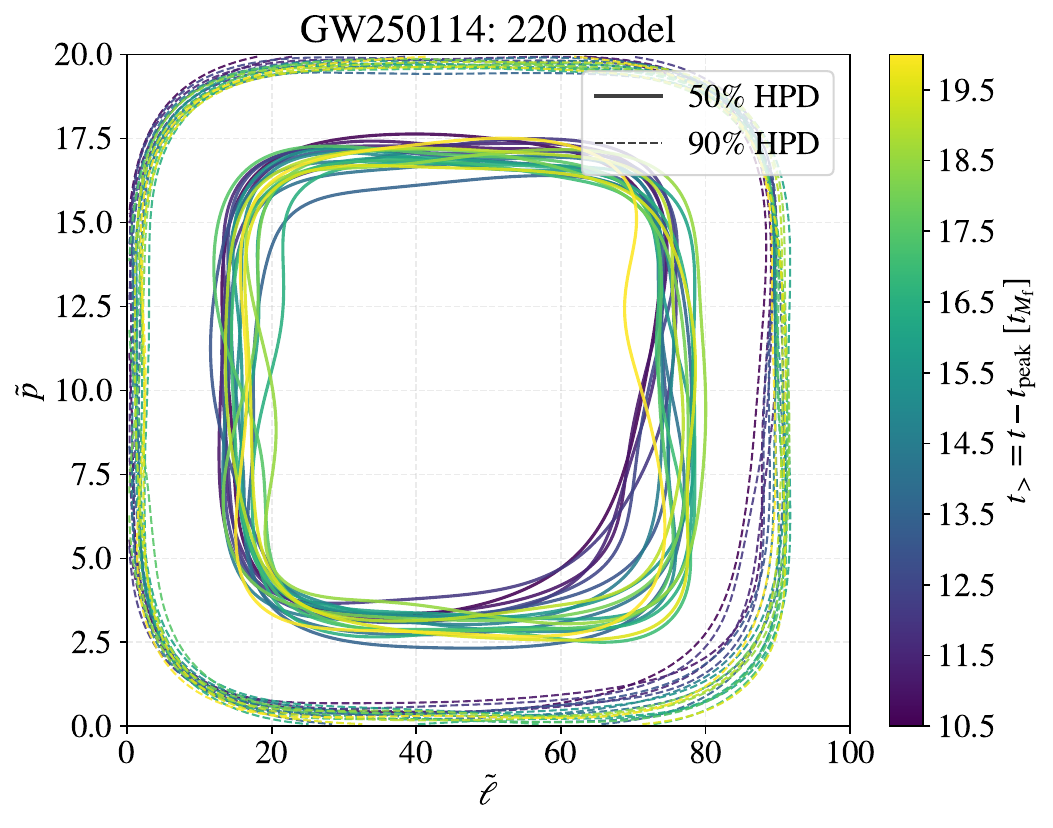}
    \includegraphics[width=0.48\linewidth]{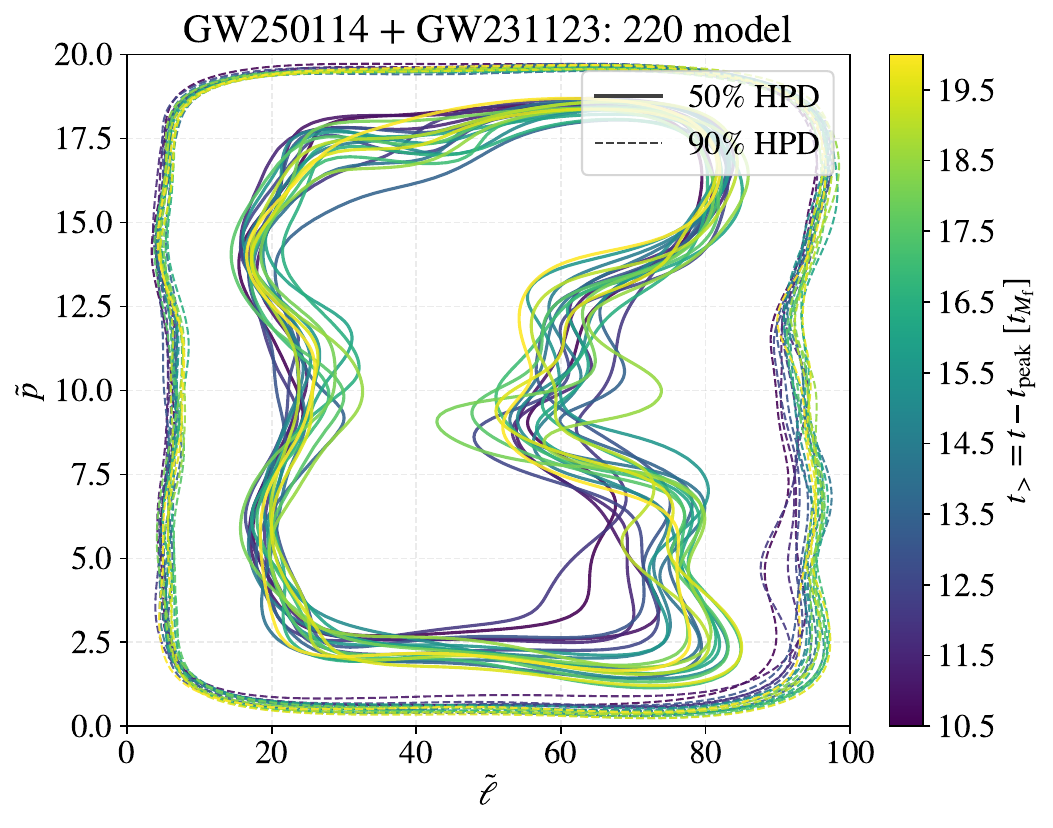}
    \caption{Two-dimensional highest posterior density (HPD) regions for GW250114 alone (left) and the joint analysis of GW250114 and GW231123 (right), using the 220 model. Solid and dashed curves indicate the $50\%$ and $90\%$ HPD contours, respectively. Colors denote the time offset relative to the peak, in units of $t_M$.}
    \label{fig: ell_p_joint_analysis}
\end{figure}

We further add the two-dimensional posterior distribution in the $\ell-p$ plane for the 220 mode, as shown in Fig. \ref{fig: ell_p_joint_analysis}. The left panel shows the result obtained from GW250114 alone, whereas the right panel shows the joint result from GW250114 and GW231123. The solid and dashed contours enclose the $50\%$ and $90\%$ highest-posterior-density~(HPD) credible regions, respectively. Each color represents a joint constraint obtained by combining GW250114 at a different ringdown starting time with GW231123 at a fixed starting time.

The $\tilde{\ell}-\tilde{p}$ degeneracy remains visible in both cases, as indicated by the $90\%$ HPD contours. The reason lies in the fact that the SNR level and the number of events are not yet sufficient to distinguish the two parameters. Nevertheless, after GW231123 is included, the relative posterior weight for the $50\%$ HPD reshapes, showing the potential of breaking the degeneracy. This occurs because events with different remnant masses give rise to different degeneracy curves in the $\tilde{\ell}-\tilde{p}$ plane, and the joint posterior downweights regions favored by one event but less consistent with the other. The current two-event analysis does not fully break this degeneracy, but it demonstrates the potential of multi-event observations to progressively reduce it and enable separate constraints on $\tilde{\ell}$ and $\tilde{p}$. In the Appendix \ref{apx:An example for multi-event observations}, we present a proof-of-principle example of how a joint analysis of multi-event observations can help break the degeneracy between the two parameters. As the catalog of gravitational-wave events expands and higher SNR ringdown observations accumulate, the extended ParSpec framework has the potential to constrain both the characteristic length scale $\ell$ and its mass scaling $p$. This flexibility makes it a useful approach for testing general relativity and distinguishing among classes of alternative gravitational theories.

\section{Conclusion}\label{sec: Conclusion}

In this work, we extended the parameterized spin expansion coefficients~(ParSpec) formalism~\cite{Carullo:2021dui,Maselli:2019mjd} by simultaneously sampling the EFT scaling index $\tilde{p}$ and the characteristic length scale $\tilde{\ell}$. Compared with previous analyses in which $\tilde{p}$ is fixed to theory-motivated integer values, this extension allows us to probe a broader class of EFT-inspired corrections and provides a more theory-agnostic ringdown test. Our primary goal is to clarify the impact of the enlarged parameter space on the inference, and to assess the robustness of the extended framework with respect to the ringdown start time, QNMs content, and joint-event constraints.

Methodologically, we first introduced informative priors on the remnant mass $M_{\rm f}$, as well as on the luminosity $D_{\mathrm{L}}$, and examined their impact on the inferred QNM amplitudes. The results show that imposing informative priors improves the constraint on the $221$ overtone amplitude at early start times. This behavior can be understood from the structure of the \texttt{pyRing} waveform model, in which the strain amplitude is controlled by the combination $M_{\rm f}/D_{\mathrm{L}}$, reducing the allowed parameter space of $M_{\rm f}$ and $D_{\mathrm{L}}$ helps suppress degeneracies. This motivates the use of the same informative-prior strategy in the subsequent inference of $\tilde{p}$ and $\tilde{\ell}$ due to $\gamma \sim (\tilde{\ell}/M_{\rm f})^{\tilde{p}}$.

Applying this framework to GW250114, a central point of this analysis is that the condition imposed on the effective coupling $\gamma$ is not merely a prior cut, but directly shapes the geometry of the allowed $(\tilde{\ell},\tilde{p})$ parameter space. We find that conditions such as $\gamma<0.01$ or $\gamma<2$ introduce nontrivial correlations between $\tilde{p}$ and $\tilde{\ell}$, while the $\gamma<1$ condition provides the most natural choice, avoiding prior-induced bias on each other during sampling. 

In the representative condition $\gamma<1$,  we investigate the dependence of the constraints on the QNM content and ringdown starting time by calculating K-L divergence and the $90\%$ credible upper bounds $\tilde{\ell}_{90}$. Although additional QNMs can in principle provide independent spectral information and help reduce parameter degeneracies, their benefit depends on whether they are sufficiently resolved by the data. The same consideration applies to the starting time, since an earlier analysis may retain more signal but can also introduce biases from nonlinear effects and other unmodeled contributions. This suggests that the choice of starting time and mode content in the extended ParSpec framework should be matched to the SNR and the resolvability of the modes in data. Otherwise, a mismatch may occur: the parameter space is enlarged, but the data do not provide a commensurate information gain.

Finally, a key issue is the degeneracy between $\tilde{\ell}$ and $\tilde{p}$ in the extended ParSpec method. Present observations do not yet have sufficient sensitivity to constrain both parameters simultaneously. Future observations with higher ringdown SNR, especially events with resolvable overtone or higher-mode contributions, and performing hierarchical multi-event analyses, will improve the constraining power of this framework. It will also be useful to explore complementary strategies, such as fixing $\tilde{\ell}$ while sampling $\tilde{p}$. Overall, the extended ParSpec framework provides a flexible way to test generalized higher-derivative corrections and strong-field gravity with ringdown data.

\acknowledgments
This work is supported by the National Natural Science Foundation of China under Grant Nos. 12505067, 12475067, and 12235019. The author, named Jia-Ning Chen, would like to thank Prof. G. Carullo for his sincere encouragement and inspiration, and He Wang for his help in data analysis. Codes are available upon request.

\appendix
\section{The results with \texorpdfstring{$\tilde{p}<10$}{p<10} prior choice}
\label{apx:the results of the prior choice}
\begin{figure}
    \centering
    \includegraphics[width=0.8\linewidth]{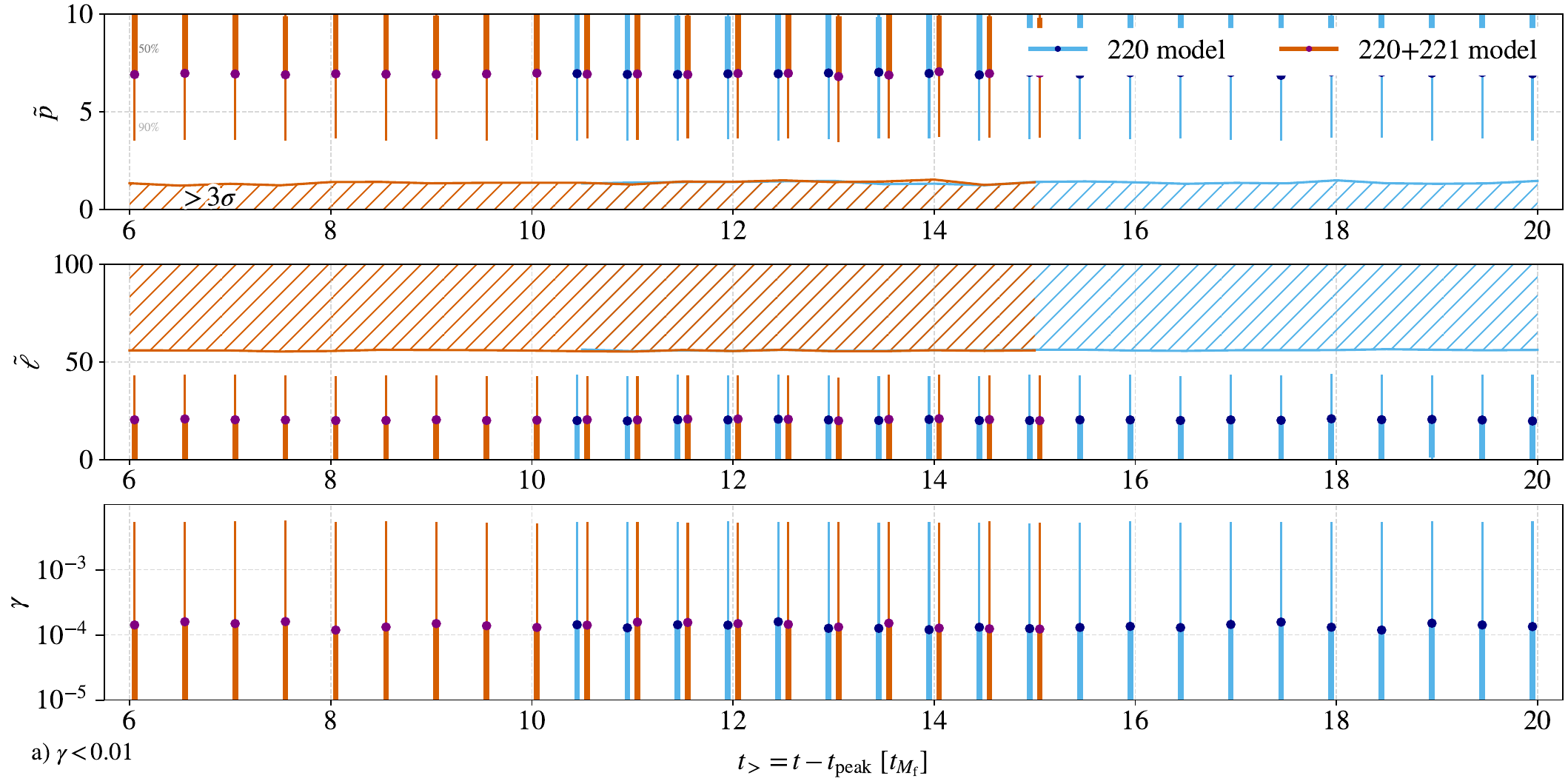}
    \includegraphics[width=0.8\linewidth]{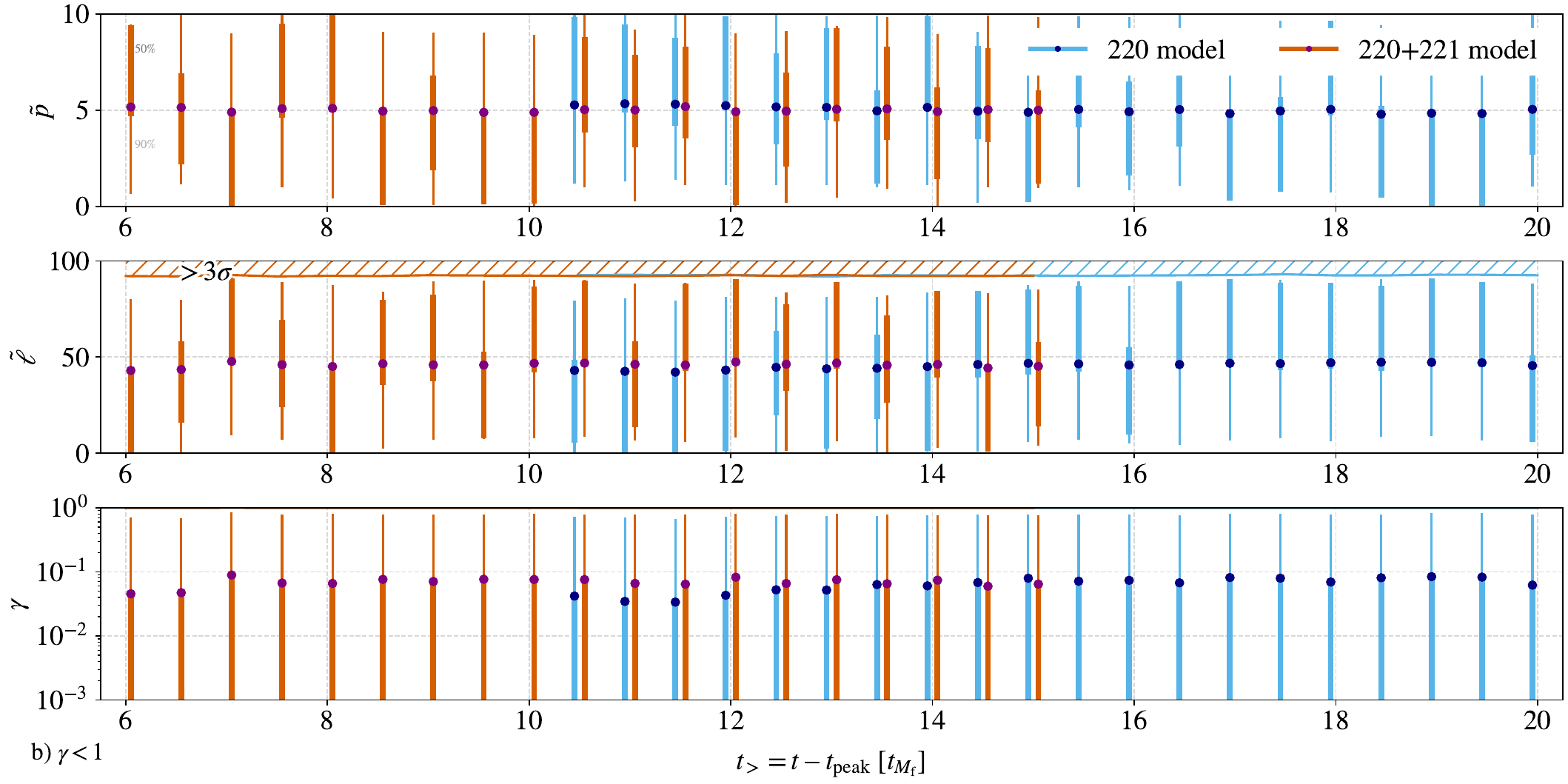}
    \includegraphics[width=0.8\linewidth]{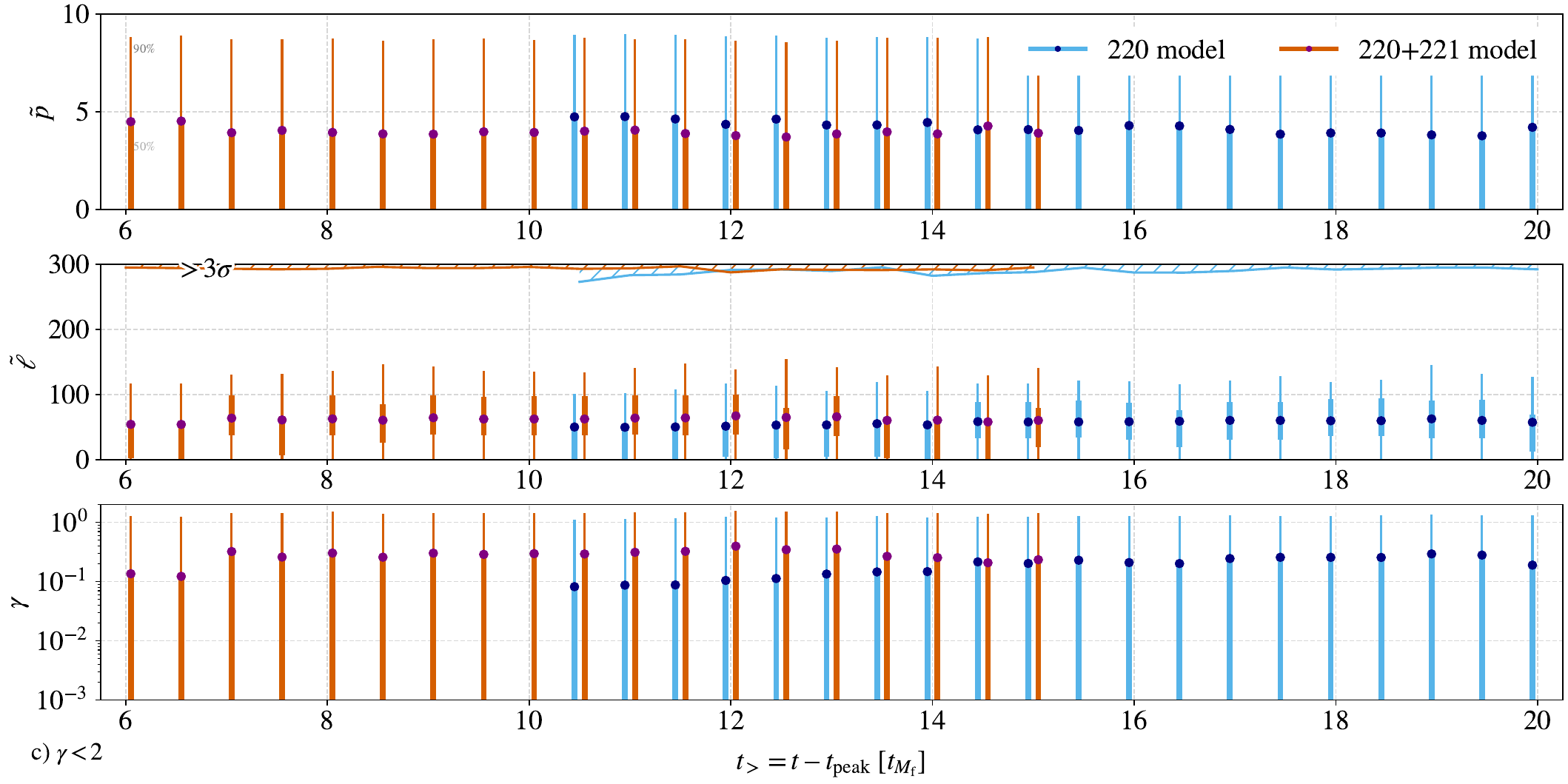}
    \caption{The posterior distributions of $\tilde{p}$, $\tilde{\ell}$, and $\gamma$ with GW250114 under $\tilde{p}<10$, provided by $\texttt{pyRing}$ package. The three pannels correspond to the exclusion conditions $a) \, \gamma<0.01$, $b) \, \gamma<1$, and $c) \, \gamma<2$. In each case, the $220$-only results are shown in light blue, and the $220+221$ results are shown in orange. In all three cases, the posterior distributions are shown as functions of the ringdown start time, and it is scanned with an interval of $0.5\,t_{M_{\mathrm{f}}}$. The black central points indicate the median of the posterior samples, while the thick and thin error bars denote the $50\%$ and $90\%$ credible intervals, respectively. The shaded region corresponds to the parameter space excluded at the $3\sigma$ confidence level.}
    \label{fig:p<10 results}
\end{figure}

\begin{figure}
    \centering
    \includegraphics[width=0.9\linewidth]{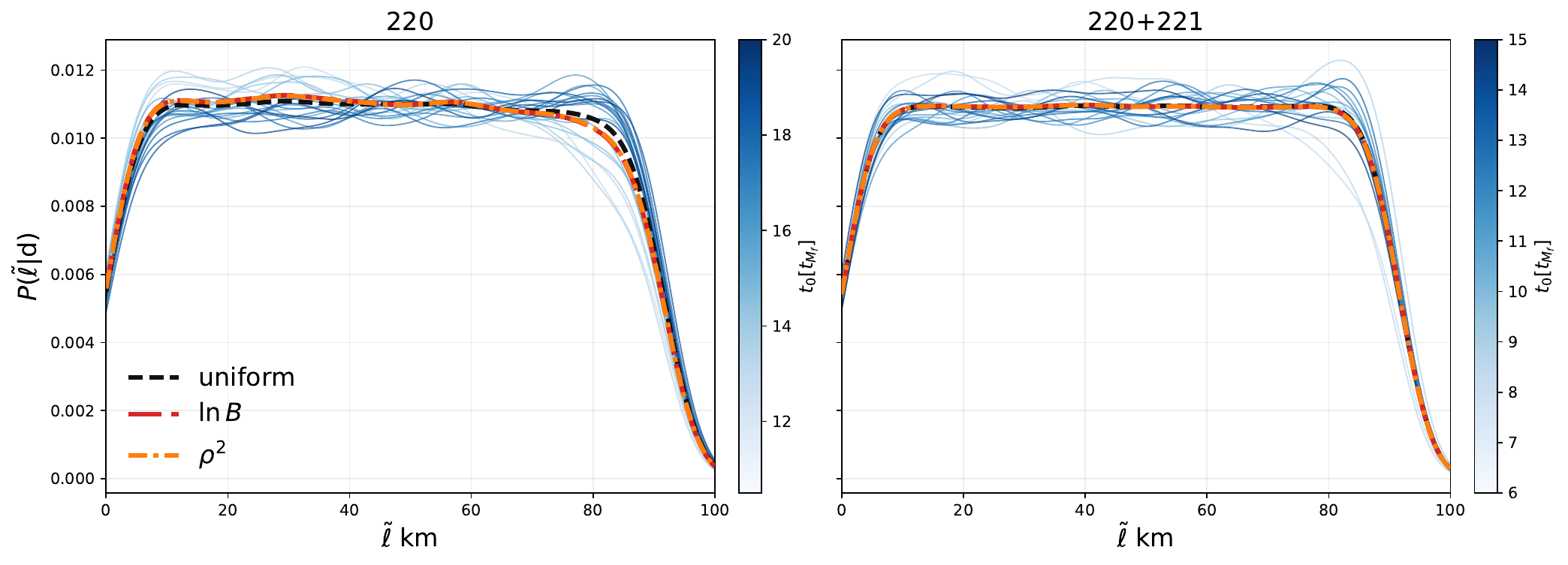}
    \caption{One-dimensional posterior distributions of the characteristic length scale $\tilde{\ell}$ with $\tilde{p}<10$ for the $220$ and $220+221$ models under the representative condition $\gamma<1$. The thin blue curves show the posteriors obtained at individual ringdown start times $t_0$, with the colorbar indicating the value of $t_0[t_{M_{\rm f}}]$. The thick curves denote the posteriors marginalized over $t_0$ using three weighting prescriptions: the black dashed curve corresponds to uniform weighting over the selected start-time range, the red dash-dotted curve corresponds to $\ln B$-based weighting, where $\ln B = \ln (Z_{\rm signal} / Z_{\rm noise})$ and the orange dash-dotted curve corresponds to $\rho^2$-based weighting, where $\rho$ is the estimated ringdown SNR.}
    \label{fig:marginalized ell in p<10}
\end{figure}

Here we present the results obtained for the prior choice $\tilde{p}<10$. The analysis setup and convention of the posterior distributions remain identical to those in the $\tilde{p}<20$ case. Compared with the $\tilde{p}<20$ case, the $\tilde{p}<10$ analysis yields a larger $3\sigma$ exclusion region in the $\tilde{\ell}$ posterior under the $\gamma<0.01$ condition, and a smaller $3\sigma$ exclusion region in the $\tilde{\ell}$ posterior under the $\gamma<2$ condition, as shown in panel a) and c) of Fig. \ref{fig:p<10 results}, respectively. The $\tilde{p}$, $\tilde{\ell}$ and $\gamma$ posteriors remain similary with those in $\tilde{p}<20$ case.  This feature is consistent with the parameter space structure, as shown in Fig. \ref{fig:parameter space}, where the condition deviated from $\gamma<1$ reshapes the allowed region in the $(\tilde{\ell},\tilde{p})$ plane. We further describe the one-dimensional posterior distributions of the characteristic length scale $\tilde{\ell}$ with $\tilde{p}<10$ for the $220$ and $220+221$ models under the representative condition $\gamma<1$, as shown in Fig. \ref{fig:marginalized ell in p<10}.

\section{The results incorporating GW231123} \label{appendix:GW231123}
In this appendix, we present the posterior results obtained from GW231123 and investigate the corresponding joint-event constraints on $\tilde{\ell}$. Compared with GW250114, GW231123 has a larger remnant mass, which provides a useful complementary mass scale for examining the behavior of the $(\tilde{\ell},\tilde{p})$ parameter space. Given the moderate SNR and the complicated start-time analysis of GW231123, we focus on its role in the joint constraint rather than on a full start-time-dependent analysis. We therefore fix the start time to $t_0=41 \, \mathrm{ms}$ for the $220$-only model and $t_0=32.3 \,\mathrm{ms}$ for the $220+221$ model after the peak of strain. The power spectral density is estimated following the procedure adopted in the official analysis, as well as other settings.

\begin{figure}
    \centering
    \includegraphics[width=0.85\linewidth]{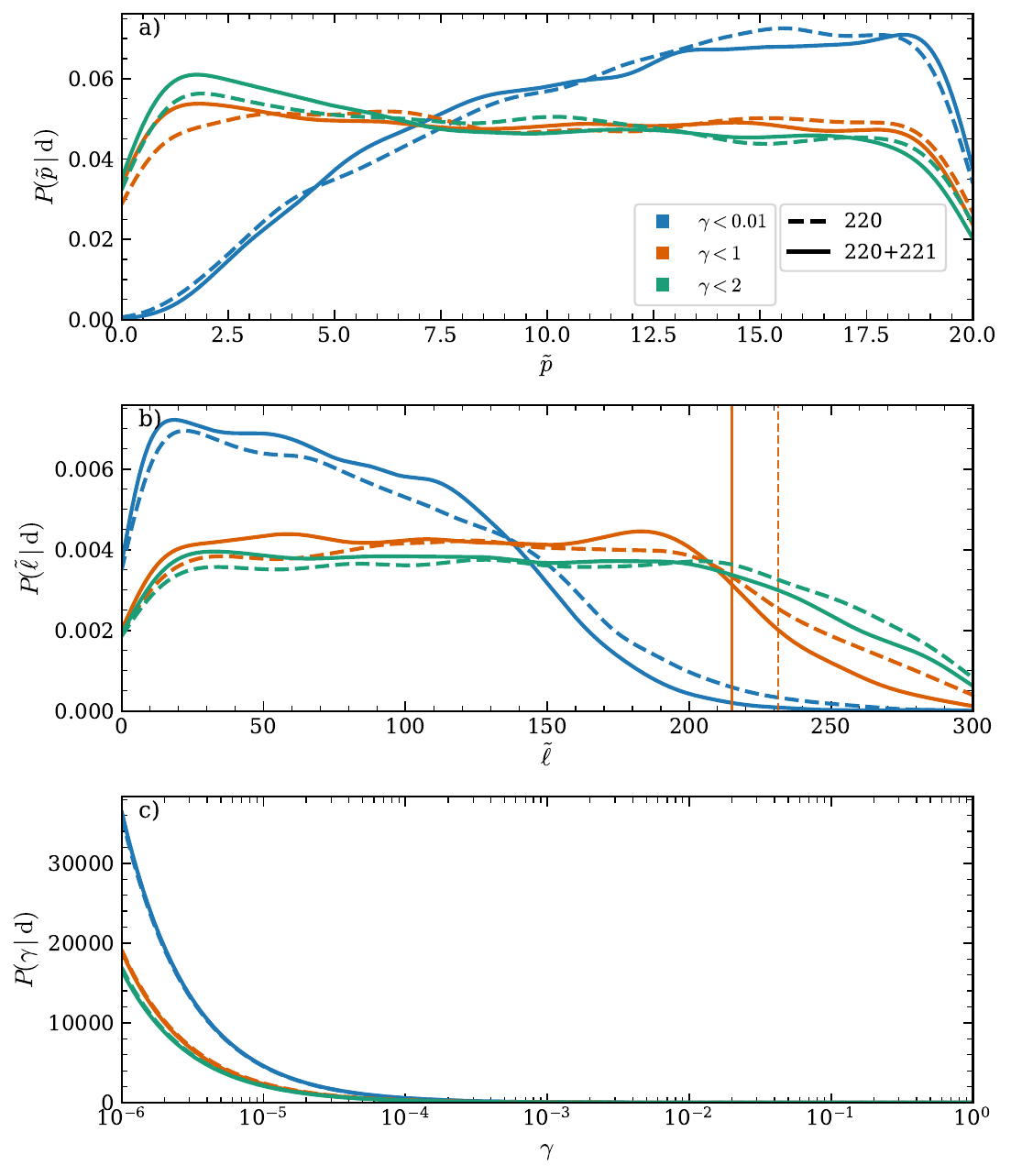}
    \caption{
    Marginalized posterior distributions of $\tilde{p}$, $\tilde{\ell}$, and $\gamma$ obtained from GW231123 under $\tilde{p}<20$. Dashed and solid curves correspond to the $220$ and $220+221$ models, respectively. Different colors denote the exclusion prescriptions $\gamma<0.01$, $\gamma<1$, and $\gamma<2$. Similar to the GW250114 analysis, the posterior distribution of $\tilde{p}$ remains weakly constrained, while the posterior distributions of $\gamma$ favor the small-coupling regime.
    }
    \label{fig:gw231123_posteriors}
\end{figure}

Fig. \ref{fig:gw231123_posteriors} shows the marginalized posterior distributions of $\tilde{p}$, $\tilde{\ell}$, and $\gamma$ obtained from GW231123 under $\tilde{p}<20$. The dashed and solid curves correspond to the $220$ and $220+221$ models, respectively, while different colors denote the different prescriptions $\gamma<0.01$ (blue), $\gamma<1$ (orange), and $\gamma<2$ (green). Similar to the GW250114 results, the posterior distribution of $\tilde{p}$ remains weakly constrained and prior-dominant, indicating that the current ringdown data still lack sufficient sensitivity to distinguish different EFT scaling behaviors. The posterior distributions of $\gamma$ again favor the small-coupling regime, showing no statistically significant evidence for deviations from the Kerr spectra. For the fundamental length scale $\tilde{\ell}$, the inferred posterior distributions exhibit the same qualitative dependence on the $\gamma$ prescriptions as observed in GW250114. In the $\gamma<0.01$ case, the posterior is concentrated toward relatively small values of $\tilde{\ell}$, by contrast, the weaker condition $\gamma<2$ allows a broader tail extending to larger $\tilde{\ell}$ values. The representative condition $\gamma<1$ also produces a relatively broad plateau structure. The overall similarities between GW231123 and GW250114 indicate that the qualitative features of the parameter space are relatively stable against changes in the source properties.

\section{An example for multi-event observations} \label{apx:An example for multi-event observations}

To illustrate more directly how this improvement arises, we add an idealized multi-event example here. In this example, we assume that $\tilde{\ell}$ and $\tilde{p}$ are universal parameters shared by all events, while each event has a different remnant mass. We choose the injected values
\begin{eqnarray}
    \tilde{\ell}_{\rm true}=45~{\rm km}, \quad \tilde{p}_{\rm true}=8\, ,
\end{eqnarray}
and consider six events with remnant masses in the source frame
\begin{eqnarray}
    M_{\rm f}^{\rm s} =  [90.0, 120.0, 140.0, 170.0, 200.0, 230.0] M_{\odot}\, .
\end{eqnarray}
All masses are converted consistently into geometric units. For each event, the injected deformation parameter is calculated as
\begin{eqnarray}
    \gamma_i^{\rm true} =\Big(\frac{\tilde{\ell}_{\rm true}}{M_{\mathrm{f},i}^{\rm s}}\Big)^{\tilde{p}_{\rm true}}\, .
\end{eqnarray}
To handle the wide range of $\gamma_{i}^{\rm true}$ values across different events and improve numerical stability, we use 
\begin{eqnarray}
    \ln\gamma_i^{\rm true}=\tilde{p}_{\rm true}\ln\Big(\frac{\tilde{\ell}_{\rm true}}{M_{\mathrm{f},i}^{\rm s}}\Big)\, .
\end{eqnarray}
throughout the subsequent analysis. 

When $\ln\gamma^{\rm true}$ is inferred from gravitational-wave data, detector noise prevents the exact recovery of its true value and may lead to some deviation or uncertainty. To mimic this effect in our illustrative analysis, we add a noise realization to $\ln\gamma^{\rm true}_{i}$ and define the resulting quantity as the mock observed value $\ln\gamma^{\rm obs}_{i}$
\begin{eqnarray}
    \ln \gamma_{i}^{\rm obs}=\ln \gamma_i^{\rm true} + n_i\, ,\quad n_i\sim\mathcal{N}(0,\sigma_{i}^2),
\end{eqnarray}
and correspondingly,
\begin{eqnarray}
    \gamma_i^{\rm obs}=\gamma_i^{\rm true}\exp(n_i{})\, .
\end{eqnarray}
Since different events may have different signal-to-noise ratios, we allow their noise levels to vary. Specifically, we use Python's \texttt{random} module to draw $\sigma_i$ uniformly from $[4,8]$ to illustrate a regime in which individual measurements of $\ln\gamma_{i}^{\rm true}$ are weak and have different precisions. Once $\sigma_i$ is fixed for each event, we draw a noise realization $n_i$ from $\mathcal{N}(0,\sigma_i^2)$. The overall simulation values used in this example are listed in Tab. \ref{tab: mock_observations}.

\begin{table*}[t]
\centering
\caption{Injected values and mock measurements for six events.}
\label{tab: mock_observations}
\small
\begin{tabular}{cccccccc}
\toprule
Event &
$M_i\,[M_\odot]$ &
$\sigma_i$ &
$n_i$ &
$\ln\gamma_i^{\rm true}$ &
$\ln\gamma_i^{\rm obs}$ &
$\gamma_i^{\rm true}$ &
$\gamma_i^{\rm obs}$ \\
\midrule
E1 &  90.0 & 4.592816 &  1.349705 &  -8.663492 &  -7.313787 & $1.727799\times10^{-4}$ & $6.662889\times10^{-4}$ \\
E2 & 120.0 & 5.327958 & -6.827831 & -10.964949 & -17.792780 & $1.729750\times10^{-5}$ & $1.873673\times10^{-8}$ \\
E3 & 140.0 & 6.010615 &  0.220989 & -12.198154 & -11.977165 & $5.039749\times10^{-6}$ & $6.286127\times10^{-6}$ \\
E4 & 170.0 & 5.309879 & -2.934951 & -13.751402 & -16.686353 & $1.066208\times10^{-6}$ & $5.665115\times10^{-8}$ \\
E5 & 200.0 & 4.674896 &  0.467964 & -15.051554 & -14.583590 & $2.905315\times10^{-7}$ & $4.639031\times10^{-7}$ \\
E6 & 230.0 & 6.664892 &  4.994049 & -16.169649 & -11.175601 & $9.497527\times10^{-8}$ & $1.401194\times10^{-5}$ \\
\bottomrule
\end{tabular}
\end{table*}

We adopt uniform priors on $\tilde{\ell}\in[0,300]~{\rm km}$ and $\tilde{p}\in[0,20]$. At each grid point, the predicted value for the $i$-th event is
\begin{eqnarray}
    \ln\gamma_i^{\rm grid}(\tilde{\ell},\tilde{p})= \tilde{p} \ln \Big(\frac{\tilde{\ell}}{M_{\mathrm{f},i}^{\rm s}}\Big)\, ,
\end{eqnarray}
Then, we evaluate the corresponding Gaussian log-likelihood for every single event
\begin{eqnarray}
    \ln \mathcal{L}_i(\tilde{\ell},\tilde{p})=-\frac{1}{2}\frac{\Big(\ln\gamma_i^{\rm grid}-\ln\gamma_i^{\rm obs}\Big)^2}{\sigma_i^2}+{\rm const}\, .
\end{eqnarray}
For multiple independent events sharing the same $\tilde{\ell}$ and $\tilde{p}$, the joint log-likelihood is obtained by multiplying the individual likelihoods
\begin{eqnarray}
    \ln\mathcal{L}_{\rm joint}(\tilde{\ell},\tilde{p})=\sum_{i=1}^{6}\ln\mathcal{L}_i(\tilde{\ell},\tilde{p})\, .
\end{eqnarray}
The joint posterior is then
\begin{eqnarray}
    P(\tilde{\ell},\tilde{p}\mid\{d_i\})\propto\pi(\tilde{\ell},\tilde{p})\prod_{i=1}^{N}\mathcal{L}_i(\tilde{\ell},\tilde{p})\, ,
\end{eqnarray}
where $\pi(\tilde{\ell},\tilde{p})$ is the common prior on the universal parameters. We use this posterior to construct the $50\%$ and $90\%$ credible regions.

\begin{figure}
    \centering
    \includegraphics[width=0.9\linewidth]{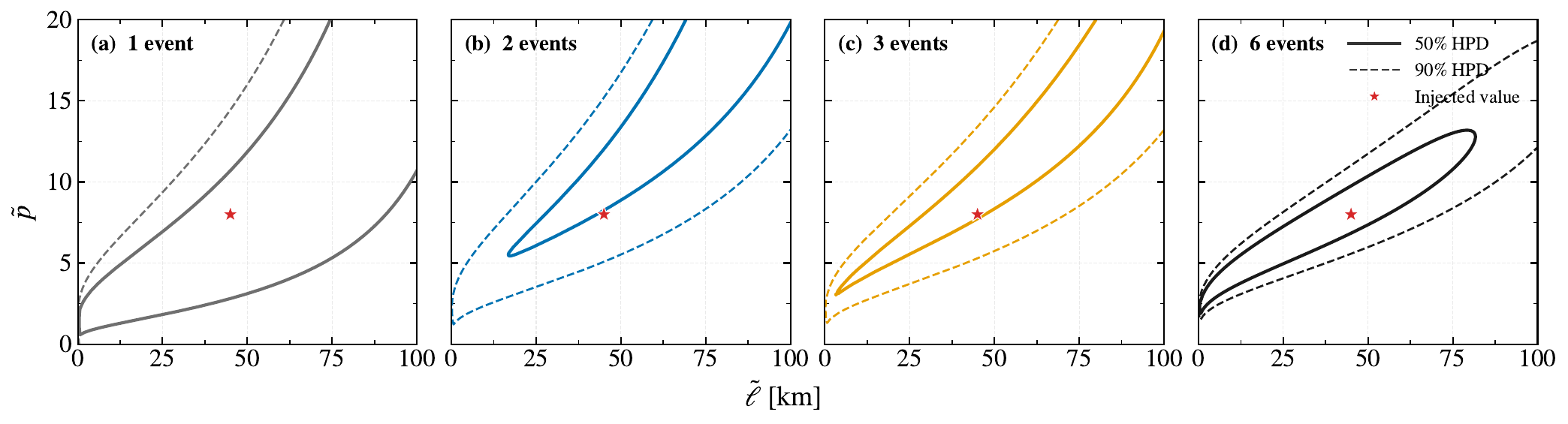}
    \caption{Figure X. Highest posterior density (HPD) contours obtained from analyses of (a) one, (b) two, (c) three, and (d) six events. Solid and dashed curves denote the $50\%$ and $90\%$ HPD contours, respectively, while the red star marks the injected parameter values.}
    \label{fig: multipile events joint observations}
\end{figure}

As shown in Fig. \ref{fig: multipile events joint observations}, the first panel shows the posterior from one event at the adopted noise level, while the subsequent panels show the results of successively combining events with different masses. The rightmost panel displays the joint constraint from all six events. As the number of events increases, the overlap between their individual degeneracy directions becomes progressively smaller, and the posterior concentrates increasingly around the injected values of $\tilde{\ell}$ and $\tilde{p}$.

We emphasize that this idealized example is intended here to illustrate how multi-events with different remnant masses can jointly constrain the common parameters $\tilde{\ell}$ and $\tilde{p}$. In a full injection-and-recovery analysis, \texttt{pyRing} will compare the injected data with waveforms generated from a given parameter space by calculating the likelihood. Here, we instead construct mock measurements of $\ln\gamma_i$ and evaluate their likelihoods directly, and the resulting constraints should therefore be interpreted as a demonstration of the multi-event mechanism. Such a full injection-and-recovery analysis would require a substantially more extensive study, accounting for solving the QNM frequencies, constructing the injected waveform, estimating the noise level, analysing the posterior parameters, and so on. 

\bibliographystyle{JHEP}
\bibliography{biblio.bib}

\end{document}